\documentclass[12pt]{article}
\usepackage{amssymb}
\usepackage{amsmath}

\usepackage{epsf}
\newcommand{\myfig}[3]{
\begin{figure}[ht] \begin{center}      \leavevmode
\epsfxsize=#2cm        \epsfbox{#1}    \end{center}
\caption{#3}   \label{fig:#1}  \end{figure}}

\newcommand{\beqn}{\begin{equation}}
\newcommand{\eeqn}{\end{equation}}
\newcommand{\bea}{\begin{eqnarray}}
\newcommand{\eea}{\end{eqnarray}}
\newcommand{\pat}{\partial}
\newcommand{\patb}{\bar{\partial}}

\newcommand{\Dcal}{{\cal D}}

\newcommand{\Hcal}{{\cal H}}

\newcommand{\Lcal}{{\cal L}}
\newcommand{\vex}{\vec{x}}
\newcommand{\vek}{\vec{k}}

\newcommand{\half}{\frac{1}{2}}

\newcommand{\wbar}{\bar{w}}

\newcommand{\tr}{{\rm tr}}

\newcommand{\bra}{\langle}
\newcommand{\ket}{\rangle}

\newcommand{\Rmath}{\mathbb{R}}
\newcommand{\Zmath}{\mathbb{Z}}
\newcommand{\Xpar}{X_{o}}
\newcommand{\Xperp}{X_u}
\newcommand{\orbid}{\Rmath^{1,d}/\Zmath_2}
\newcommand{\atilde}{\tilde{a}}
\newcommand{\omegak}{\omega_{\vek}}

\newcommand{\comment}[1]{}

\begin{document}

\begin{titlepage}
\begin{flushleft}
       \hfill                      {\tt hep-th/0304241}\\
       \hfill                      HIP-2003-28/TH \\
       \hfill                      CERN-TH/2003-097 \\
       \hfill                      November 20, 2003\\
\end{flushleft}
\vspace*{3mm}
\begin{center}
{\Large {\bf The Taming of Closed Time-like Curves}}
\end{center}
\begin{center}
\vspace*{12mm}

\renewcommand{\thefootnote}{\fnsymbol{footnote}}
{\large Rahul Biswas${}^{3}$\footnote{E-mail: rbiswas@students.uiuc.edu}, Esko
Keski-Vakkuri${}^{1}$\footnote{E-mail:
esko.keski-vakkuri@helsinki.fi}, Robert G.
Leigh${}^{2,3}$\footnote{E-mail: rgleigh@uiuc.edu}, \\  Sean
Nowling${}^{3}$\footnote{E-mail: nowling@students.uiuc.edu}
and Eric Sharpe${}^{3}$\footnote{E-mail: ersharpe@uiuc.edu}}
\vspace*{4mm}

{\em ${}^{1}$Helsinki Institute of Physics \\
P.O. Box 9,
FIN-00014  University of Helsinki,
Finland \\}
\vspace*{5mm}
{\em ${}^{2}$CERN-Theory Division\\
CH-1211, Geneva 23, Switzerland\\}
\vskip .5cm
{\em ${}^{3}$Department of Physics,
University of Illinois\\
1110 W. Green Street, Urbana, IL 61801, U.S.A.}
\vspace*{10mm}
\end{center}

\begin{abstract}
We consider a $\orbid$ orbifold, where $\Zmath_2$ acts by time and
space reversal, also known as the embedding space of the
elliptic de Sitter space. The background has two potentially dangerous
problems: time-nonorientability and the existence of closed
time-like curves. We first show that closed
causal curves disappear after a proper definition of the time
function. We then consider the one-loop vacuum expectation value
of the stress tensor. A naive QFT analysis yields a divergent
result. We then analyze the stress tensor in bosonic string
theory, and find the same result as if the target space would be
just the Minkowski space $\Rmath^{1,d}$, suggesting a zero result
for the superstring.
This leads us to propose a proper reformulation of QFT, and recalculate the
stress tensor. We find almost the same result as in Minkowski
space, except for a potential divergence at the initial time slice
of the orbifold, analogous to a spacelike Big Bang singularity.
Finally, we argue that it is possible to define local S-matrices,
even if the spacetime is globally time-nonorientable.
\end{abstract}

\end{titlepage}

\baselineskip16pt

\section{Introduction and Summary}
A technical obstacle in exploring string theory in time-dependent
space-times is to find suitable backgrounds where string
quantization is tractable. Early work includes
\cite{Horowitz:1991ap} -- \cite{Lawrence:1996ct}.
More recently, interest has been revitalized, motivated in part
by novel string-based cosmological scenarios (see for example
\cite{Khoury:2001bz,Seiberg:2002hr,Veneziano:2000pz}).
An obvious path to follow was
to construct such backgrounds as time-dependent orbifolds of
Minkowski space
\cite{vijayetal} --\cite{Fabinger:2002kr}
or anti-de Sitter space \cite{Elitzur:2002rt,Craps:2002ii,Simon:2002cf}.
Further related work includes
\cite{Aharony:2002cx} -- \cite{Cornalba:2003ze}.
However, depending on
how the orbifold identifications are defined, potentially dangerous issues
may arise. The resulting time-dependent orbifolds can have regions
with closed time-like curves (CTCs) or closed null curves (CNCs),
or may not even be globally time-orientable. Therefore,
one could choose to first make a list of desirable features for the
orbifolds and then try to limit the study only to those backgrounds
that possess those features.  This sensible strategy was laid out and
pursued by Liu, Moore and Seiberg \cite{Liu:2002ft,Liu:2002kb}.
For orbifolds of type $\Rmath^{1,3}/\Gamma $ where $\Gamma$
is a discrete subgroup of the Poincar\'e group, the list turned
out to be very short containing only the null branes with $R>0$.
However, the null brane construction involves identifications by arbitrarily large
boosts. This turns out to be another
potential reason for instabilities, and it was argued by Horowitz
and Polchinski \cite{Horowitz:2002mw} that such backgrounds become
unstable after just a single particle is added, because on the
covering space the particle can approach its infinitely many
images with increasingly high momenta and produce a black hole.
Additional discussion of potential problems can be found in \cite{Lawrence:2002aj,
Liu:2002kb,Berkooz:2002je}.

Aside from constructing and studying time-dependent backgrounds
by alternative methods, one might speculate if the
list of desirable features for
suitable orbifold backgrounds was too prohibitive and reconsider the
reasons for including each item on the list. In any case, it is important to understand
if and/or why string theory actually has problems with these features. The reason for
demanding that there be no regions containing closed time-like
curves appears obvious. Classically, CTCs violate causality, and quantum mechanically, coherence and unitarity come into question. It has been
conjectured by Hawking \cite{Hawking:1992nk} that the laws of
physics prevent CTCs from appearing if they do not exist in the past. The arguments in support of
this chronology protection conjecture (CPC) are usually based on
general relativity plus matter at the classical or semiclassical
level. A recent summary can be found in \cite{Visser:2002ua}.
Essential features are that perturbations can keep propagating
around a CTC so that backreaction accumulates, or quantum effects
can lead the matter stress tensor to diverge at the boundary of
the CTC region, leading to infinite backreaction. However, the
trouble with CTCs and CNCs seems to arise from propagation along
them, rather than merely from their existence. It is not clear if
the two are equivalent. For example, the model studied in
\cite{vijayetal} involves CTCs and CNCs, but it was argued that
they do not necessarily pose a problem in quantum mechanics if one can
project to a subspace of states which do not time evolve along
the CTCs and CNCs. Another desirable feature on the list was
time-orientability. This was included to avoid problems in
defining an S-matrix, and problems associated with the existence of
spinors \cite{Kay,Chamblin:1995jz,Chamblin:1995nq}.
However, the consequences of a
lack of time-orientability have not yet been subject to extensive
investigation and are thus less well understood. From the point of view of
local physics, one might wonder if the whole Universe could be
globally time-nonorientable, but in such a way that the global
feature could only be detected by meta-observers and never be
revealed by local experiments. The orbifold studied in
\cite{vijayetal} is an example of a spacetime which is globally
time-nonorientable. In any case, its structure appears to allow for a definition of an S-matrix for local experiments.

To summarize, there are many reasons to investigate the chronology
protection conjecture and time-nonorientability. We also note that
recently the former topic has been investigated from other points
of view in the context of string theory and holography
\cite{Boyda:2002ba,Herdeiro:2002ft,Harmark:2003ud,Gimon:2003ms,
Dyson:2003zn}.
The $\orbid$ orbifold, obtained by identifying points $X$ with
reflected points $-X$, provides a simple model which incorporates
both issues. Some comments were made in passing in
\cite{vijayetal}. In this paper we perform a more detailed
investigation.

The orbifold is also relevant for the elliptic interpretation of
de Sitter space ($dS$)
\cite{Schrodinger1,Gibbons:1986dd,Folacci:1987gr,Parikh:2002py}. A
$d$-dimensional de Sitter space is a time-like hyperboloid
embedded in $\Rmath^{1,d}$. The $\Zmath_2$ reflection on
$\Rmath^{1,d}$ induces an antipodal reflection on the $dS$
spacetime. The elliptic de Sitter space $dS/\Zmath_2$ is then
defined by identifying the reflected antipodal points.

The identification leads to various problems in quantum field
theory.  Previous studies of the elliptic $dS$ spacetime have
discussed problems in defining a global Fock space in the global
patch; however, it was possible to construct QFT and a Fock space
by restricting to the static patches of observers at the
(identified) north and south poles. The same problem is
encountered in trying to formulate QFT on $\orbid$. Moreover,
there is a question of whether the orbifold is an unstable
background. One can present a quick semiclassical derivation of
the stress energy and find that it diverges; for example in the case of
a massless scalar field one obtains a divergence in the lightcone
emanating from the origin.

We show that these problems can be circumvented after one
formulates quantum field theory in a manner which appropriately
incorporates the $\Zmath_2$ identification under time reversal and
space reflection. We argue that in such a formulation one needs to
first double the field degrees of freedom, with the copy fields
propagating towards the reversed time direction, and then identify
the degrees of freedom under the $\Zmath_2$ reflection. The
doubling of fields is motivated by (the zero temperature limit of)
the real-time formulation of finite temperature QFT. The doubling
of degrees of freedom helps to overcome problems with causality
when the light-cones of the identified points $X$ and $-X$
intersect, as we will assume that the two copies of the fields (at
$X$ and $-X$) are dynamically decoupled. Note that in the limit in
which the cosmological constant approaches zero, the $dS$
spacetime becomes locally Minkowski spacetime. Correspondingly, it
has been argued that in this limit the elliptic $dS$ spacetime
goes to two copies of Minkowski spacetime, related by the
$\Zmath_2$ reflection \cite{Parikh:2002py}. In the present work,
we would instead propose that QFT on the elliptic $dS$ spacetime
goes to QFT on $\orbid$, with two copies of {\em fields},
identified under the $\Zmath_2$ reflection.

We also study the  backreaction at one-loop level in string
theory. We calculate the one-loop graviton tadpole in the $\orbid$
background, and show that the answer is the same as if the
background were just $\Rmath^{1,d}$! While the answer
first appears puzzling, it appears very natural in
relation to the $\Zmath_2$ invariant formulation of QFT.
Indeed, the low-energy limit of string theory should be the
$\Zmath_2$ invariant QFT.

Finally, we argue that it is possible to
define S-matrices in a manner that makes sense locally. The
definition only breaks down at the point which can be regarded as
the initial ``Big Bang singularity'' of the orbifold, and at that
point we also find that even in the invariant reformulation of the
QFT, the stress tensor diverges. However, it is also possible that
stringy effects lead to a smooth blow-up of the orbifold
singularity. Then the QFT would need to be reconsidered in this
smooth background.

We have organized the paper as follows. In Section
\ref{sec:oldorbid}, we review some features of the time-dependent
orbifold background, and focus on some novel features of these
orbifolds. In particular, we point out that a choice of time
orientation must be made. In Section \ref{sec:tradcalc}, we review
the (na\"ive) analysis of the gravitational back reaction in this
geometry. In Section \ref{sec:stringcalc}, we ask if string theory
can do better, and present similar calculations in string theory
(complementary calculations in a different formalism are shown in
the Appendix.). We find that the result differs significantly from
the na\"ive QFT analysis. To resolve the puzzle, in Section
\ref{sec:qftnew}, we present a proper formulation of quantum field
theory on the $\orbid$ background. We show that the result now
contains the familiar short-distance Minkowski spacetime
divergence, in agreement with the string calculation, plus an
additional divergence, which can be interpreted as a
``cosmological initial condition.'' The latter does not arise from
the first quantized string calculation and would need a more
involved analysis to understand as a low energy limit of string
theory. Finally, in Section \ref{sec:smat}, we discuss further
features of the interacting QFT, including a discussion of the
S-matrix.

\section{Overview of $\orbid$}
\label{sec:oldorbid}

Let us first review some features of the $\orbid$ orbifold
\cite{vijayetal}.
We begin with the covering space $\Rmath^{1,d}$ and identify the
time and space coordinates under the reflection
\begin{equation}
\label{eq:orbidaction}
(t,x^a) \sim (-t,-x^a) \ .
\end{equation}
The resulting orbifold is a space-time cone, depicted in Figure
\ref{fig:newfig1.eps} for $d=1$.
Points in the opposite quadrants (I
and III, and II and IV) are identified.
\myfig{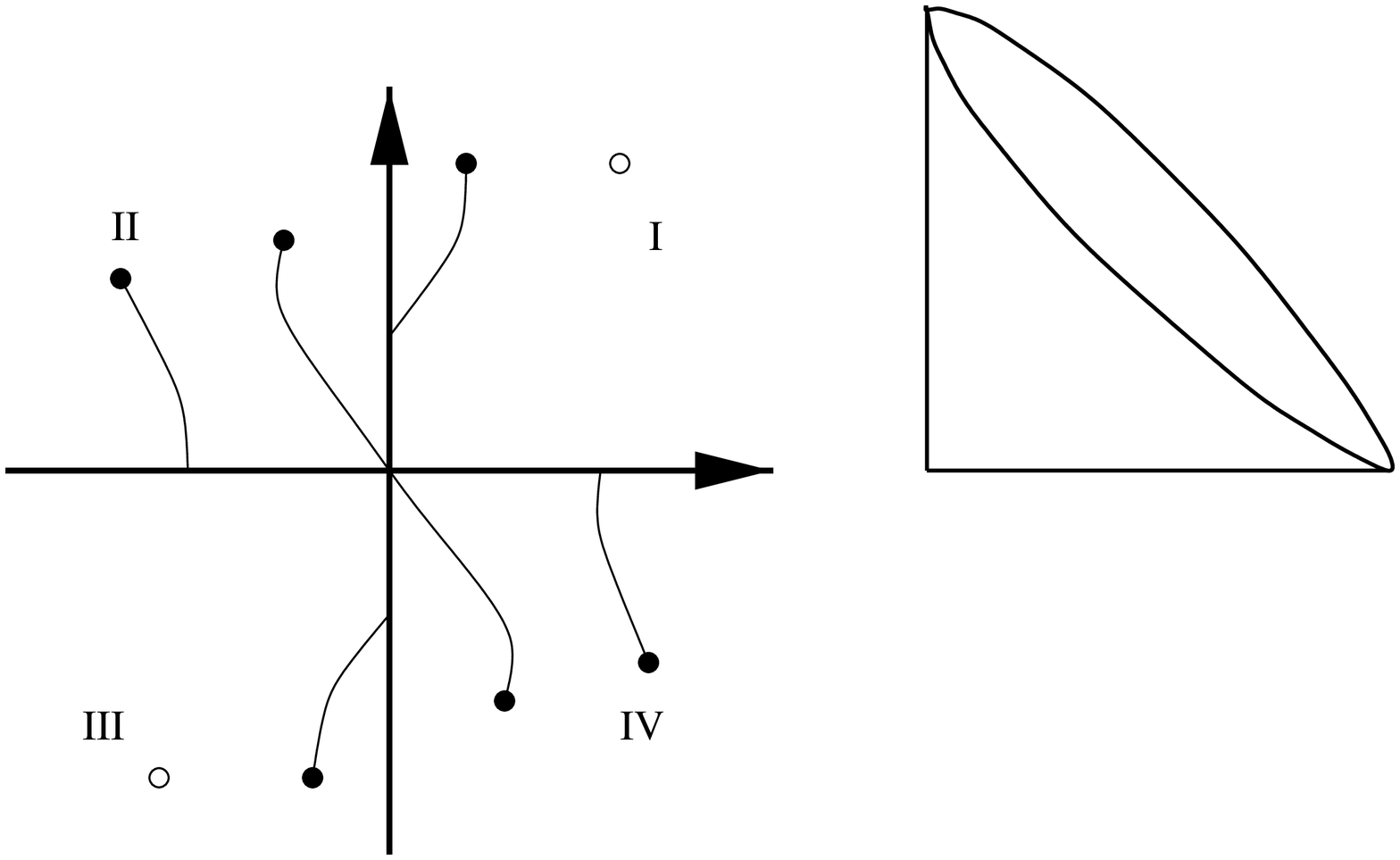}{15}{The orbifold $\Rmath^{1,1}/\Zmath_2$. Also depicted
are some identified points and resulting closed timelike curves.}
Orbifolds that act purely spatially are familiar and are certainly well
understood. New problems arise when the identification involves
the time direction; for example it is not guaranteed that the string
spectrum will be free from tachyons and ghosts.
Ref. \cite{vijayetal} investigated bosonic and type II
superstrings on $\orbid \times \Rmath^{n}$, with $n$ additional
spacelike directions added to bring the total spacetime dimension
to 26 or 10. It was shown, using a Euclidean continuation, that although the background is
time-dependent and quantization had to be done in the covariant
gauge, the physical spectrum did not contain any negative norm
states (ghosts), at least in a range of $d$. The superstring spectrum did not contain any
tachyons and the one-loop partition function vanishes for the
superstring.

Although string theory passed the first tests, questions
associated with the time identification on the orbifold remained.
In the orbifold (\ref{eq:orbidaction}), there is actually
extra data that must be specified. To see this, we note that to specify a
Lorentzian metric on an orientable space $M$ ($w_1(M)=0$), we must
specify a {\it time orientation}. Mathematically, this implies a real
rank 1 subbundle $L\subset TM$, the time orientation bundle. ($M$ is said to be
globally time-orientable when $w_1(L)=0$.) A time-like Killing vector defining time's arrow, if available, would be  a
global section of this line bundle.

In the case of the orbifold (\ref{eq:orbidaction}), we must ask how various
quantities descend from the covering space to the orbifold. In particular,
$\partial/\partial t$ is manifestly not invariant under
the group action, and so does not define a time's arrow, or
time-like Killing vector, in the quotient.  Thus, this orbifold
leaves ambiguous the direction on which time flows in the quotient
 -- we must manually make a choice of direction of time-flow.

Furthermore, the natural time orientation bundle on the covering space does
descend to the quotient space, but (omitting the singularity at the origin)
the class $w_1(L)$ is non-trivial. Thus the image of $L$ on the quotient is
not time-orientable. Although locally we can choose a perfectly sensible
notion of time orientation, this is not possible globally.

\myfig{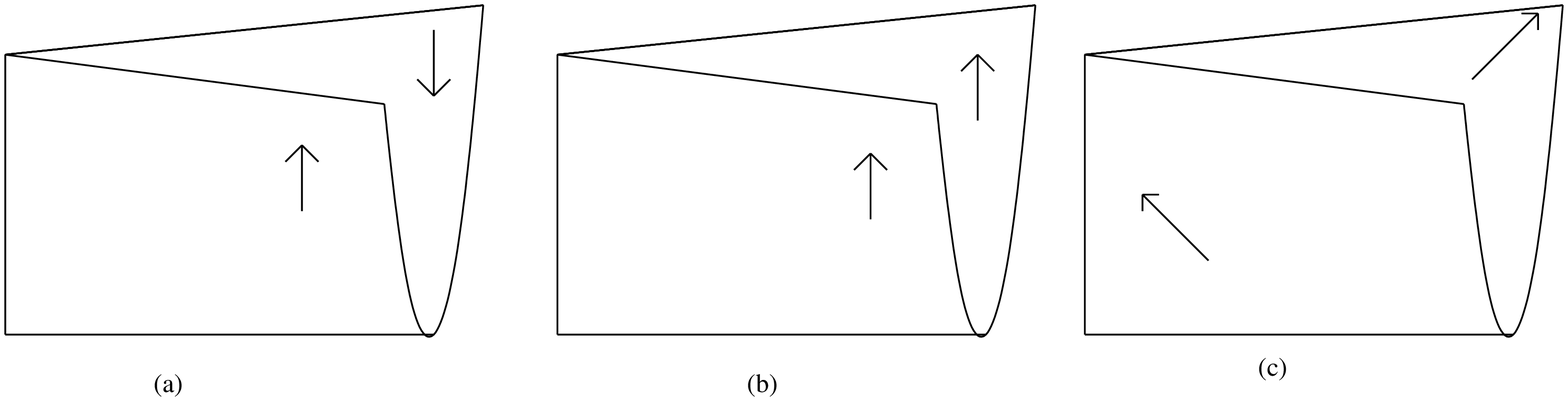}{12}{Three possible time-arrows
on the quotient $\Rmath^{1,1}/\Zmath_2$.}
To illustrate, let us consider the case of $\Rmath^{(1,1)}/\Zmath_2$.
The obvious choice of time's arrow on the
covering space $\Rmath^{1,1}$,
namely $\partial/ \partial t$, is not invariant under the group action,
a property which manifests itself in the observation that by picking different
fundamental domains for the group action on the cover, the time's arrow
in those fundamental domains restricts to a different time's arrow on the
quotient.

In Fig. \ref{fig:pic1.eps} we have shown three possible time-arrows that one
can construct on $\Rmath^{1,1}/\Zmath_2$.
The left-most case corresponds to taking the fundamental domain
to be regions I and IV, the middle case corresponds to taking the
fundamental domain to be regions I and II, and the right-most case
corresponds to taking the fundamental domain to be one
side of a wall of the lightcone through the origin.
In each case, omitting the origin, the time-orientation line bundle
on the quotient is not orientable ($w_1(L) \neq 0$),
hence each choice of time's arrow depicted in figure~\ref{fig:pic1.eps} has
zeroes -- in case (a), along the left vertical crease,
and in case (b), along the bottom horizontal crease. Note that in
each case it would also be possible to choose a reverse time
orientation (reversed arrows). Then {\em e.g.} Fig.
\ref{fig:pic1.eps}(b) would depict a ``big crunch" rather than a ``big bang."

In Figure \ref{fig:figgg3.ps}, we have drawn the quotient space
corresponding to Fig. \ref{fig:pic1.eps}(a). In this case, there are
asymptotic regions for both $t\to\pm\infty$. However, there is a topology change of constant $t$ slices at $t=0$.
\myfig{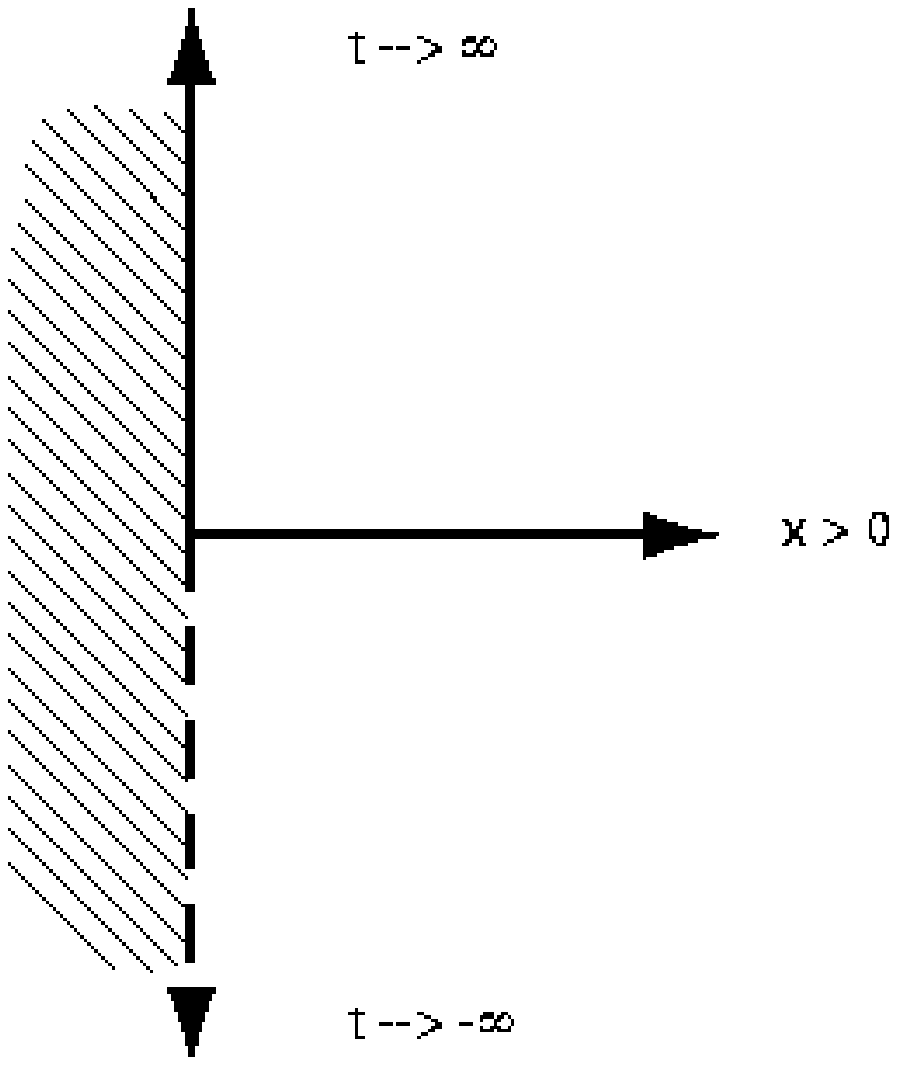}{4}{A view of the quotient spacetime (for 1+1 dimensions).
Note the absence of the $x=0$ axis for $t<0$.}
Another choice for the quotient space, corresponding to
Fig. \ref{fig:pic1.eps}(b),  is shown in Fig. \ref{fig:figg4.ps}. In this
case, there is no is no asymptotic region corresponding to $t\to -\infty$.
\myfig{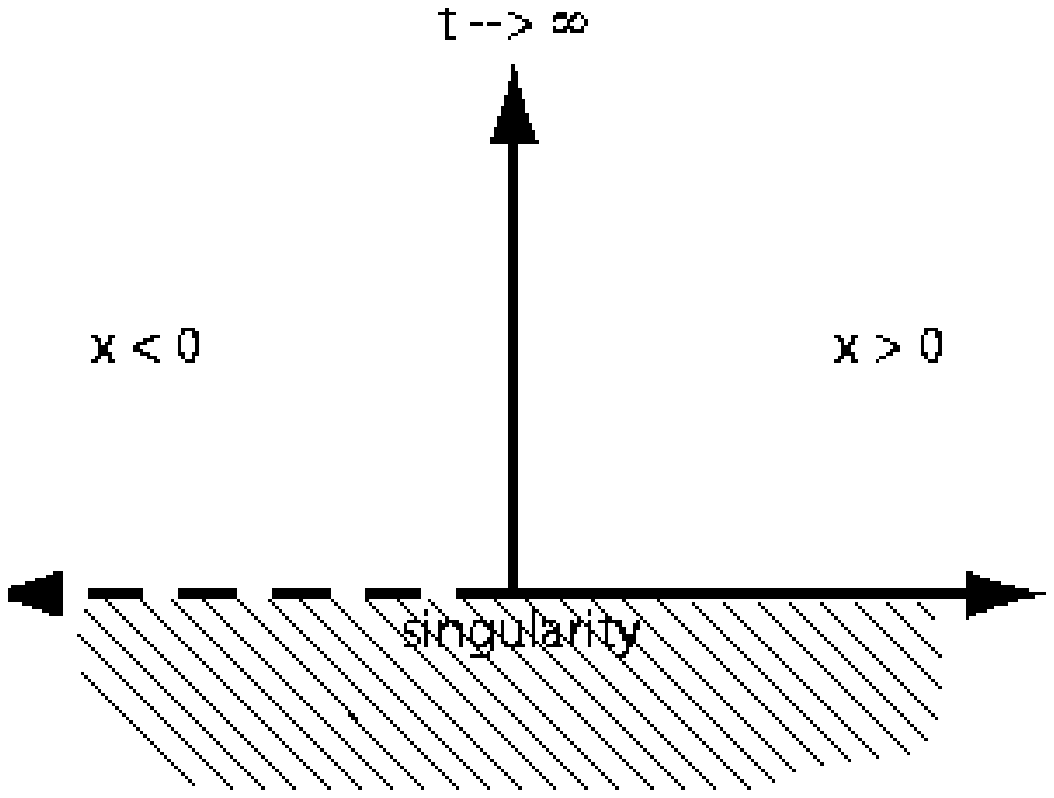}{5}{Another view of the quotient
spacetime (for 1+1 dimensions).
Note the absence of the $t=0$ axis for $x<0$. The $t=0$ axis represents
a ``big bang'' singularity--the beginning of the spacetime.}
Instead, we have a ``big bang'' singularity at $t=0$. It is interesting to
contemplate the properties of quantum field theory on such a spacetime.
It is of even more interest to ponder the role of string theory.
We will return to a more thorough discussion of these issues in a later
section.

Let us also discuss the closed time-like curves in this geometry.
In the covering space, with the natural choice of Minkowski time
orientation, there are non-trivial forward oriented closed time-like
curves. Examples are shown in
Fig. \ref{fig:newfig1.eps}.
It is clear from
the figure that there are CTC's which begin at any spacetime point.

Consider however these curves in the quotient space (let us refer
to the choice of time-orientation in Fig. \ref{fig:pic1.eps}(a) to be
definite). In going to the quotient we make a choice of (local) time
orientation which is not compatible with the time orientation of the
covering space. As a result (and this is true for any choice), the CTC's
that we identified in the covering space are {\em not forward oriented}
in the quotient. The examples given in
Fig. \ref{fig:newfig1.eps} are
redrawn in the quotient in
Fig. \ref{fig:newfig5.eps}.
\myfig{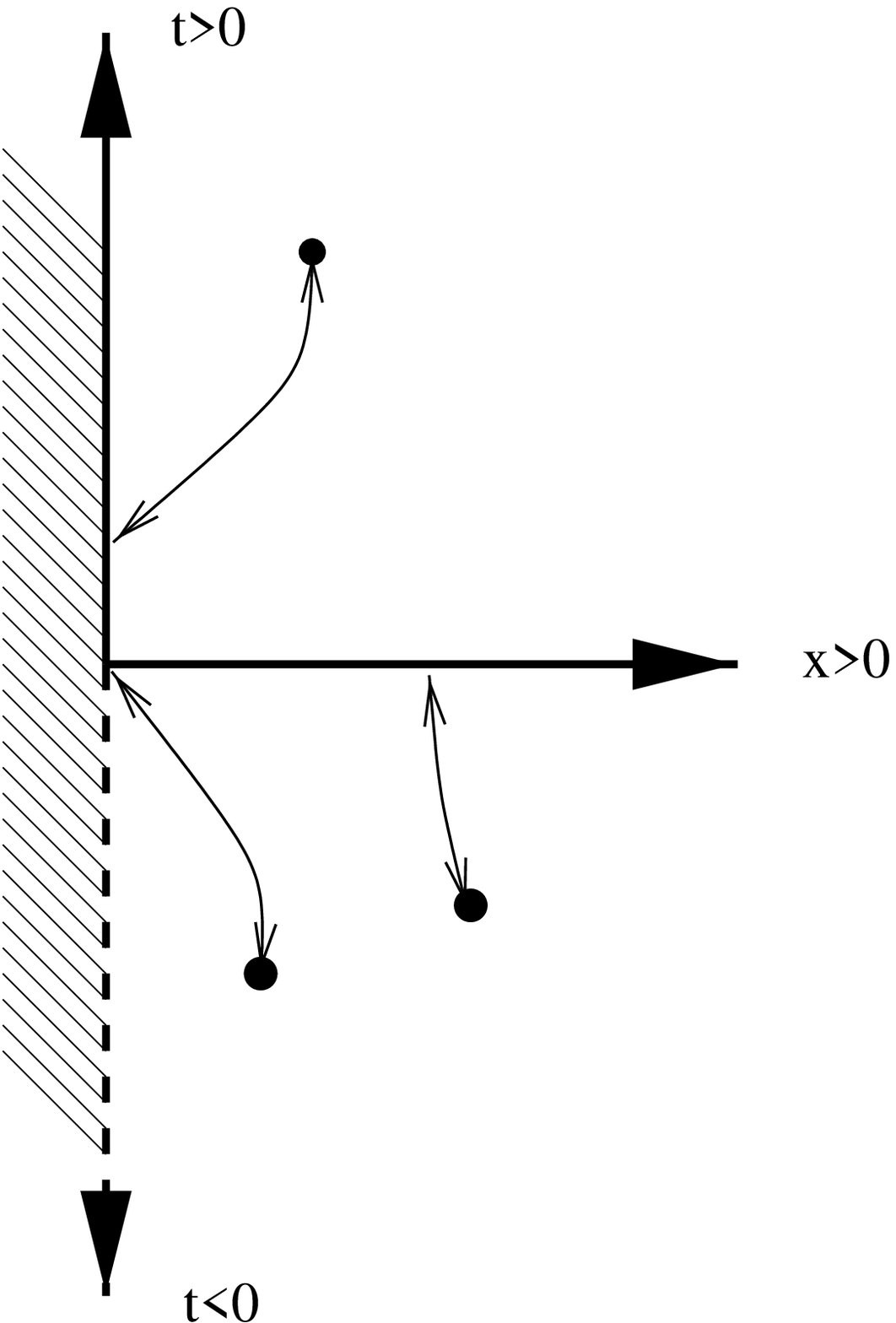}{4}{The CTC's of
Fig. \ref{fig:newfig1.eps} are not
forward oriented in the quotient.}
In fact, the only CTC in the quotient
must begin and end on the singular axis (the curve can be constructed
by a limiting procedure.)

Let us quickly review this discussion. In the Lorentzian orbifold,
a choice of time orientation must be made in the quotient.\footnote{It is
not clear how this choice should be encoded in string theory.} This gives
rise to physically inequivalent spacetimes that are singular along an axis.
The singularity is associated with an undefined time orientation. Whereas
there were oriented CTC's through every point in the covering space, (almost)
all of these are not forward oriented in the quotient. Next, we will consider
quantum field theory on this background; we focus on the issue of
back-reaction.

\section{Backreaction in Quantum Field Theory}
\label{sec:tradcalc}

Here we give a short review of the standard QFT calculation
for the vacuum expectation value (vev) of the stress tensor, which
in general leads to a divergence hinting at an instability of the
background. Later, we will contrast this with a calculation
in string theory.

The gravitational
backreaction from the renormalized stress energy of a quantum field may be
evaluated semi-classically
\begin{equation}
  G_{\mu\nu} = -8\pi G_N\bra T_{\mu\nu} \ket_{ren}.
\end{equation}
Here the subscript refers to the fact that one subtracts off the usual
vacuum energy contribution ---  the curvature is well-defined if there are
no divergences other than the usual flat space short distance singularities.
In more detail \cite{Visser:2002ua}, one defines the renormalized
stress tensor starting from the two-point correlation  function
$G(x,y)$ written in Hadamard form
as a sum over geodesics $\gamma$ from $x$ to $y$.
The expectation value of the point-split stress
tensor can then be defined as
\begin{equation}
 \bra T_{\mu\nu} (x,y,\gamma_0) \ket = D_{\mu\nu}(x,y,\gamma_0)
 G(x,y) \ ,
\label{psplit}
\end{equation}
where $\gamma_0$ denotes the trivial geodesic from $x$ to $y$
which collapses to a point as $y\rightarrow x$, and
$D_{\mu\nu}(x,y,\gamma_0)$ is the second order differential
operator associated with the action of the particular field in
scrutiny. The renormalized stress energy $\bra T_{\mu\nu}(x) \ket_{ren}$
is defined by discarding
the universal divergent piece arising from the contribution of the
trivial geodesic to the Green function. That is, one replaces in
(\ref{psplit}) the Green function by the renormalized Green
function, defined
with the trivial
geodesic excluded from the sum over geodesics:
\begin{equation}
G(x,y) = \sum_{\gamma} \cdots \rightarrow G_{ren}(x,y) = \sum_{\gamma \neq
\gamma_0} \cdots \ ,
\end{equation}
and then removing the point-splitting regularization from (\ref{psplit})
by taking the limit $\lim_{y\rightarrow x}$.

Let us then consider the
$\Rmath^{1,1}/\Zmath_2$ orbifold and {\em e.g.} the stress energy of
a free massless scalar field.
The field decomposes into left- and right-movers. Let us focus on
the right-movers only. The right-moving component of the stress
tensor is
\begin{equation}
T_{uu}(u) = :\pat_u \phi (u) \pat_u \phi (u): \ ,
\end{equation}
where $u=t-x$.
To proceed as in the above, we start from the
Minkowski space two-point correlation function
\begin{equation}
G(u,u') \sim -\ln (u-u') \ ,
\end{equation}
associated with the trivial geodesic from $(u,v)$ to $(u',v')$.
On the orbifold, the points $(u,v)$ are identified with $(-u,-v)$
and $(u',v')$ identified with $(-u',-v')$. This gives arise to
three additional geodesics (Fig. \ref{fig:figgg5.ps}), so the
two point function
on the orbifold would be
\begin{equation}
G_{orb}(u,u') = G(u,u') + G(u,-u') + G(-u,u') + G(-u,-u') \ .
\label{gorb}
\end{equation}
Subtracting off the trivial universal divergence, we then obtain
the renormalized stress energy
\bea
 \bra \tilde{T}_{uu} (u) \ket_{ren}
 &=& \lim_{u'\rightarrow u} \pat_u\pat_{u'}
 \{ -\ln (u-u') -\ln (u+u')\}_{ren} \nonumber \\
 \mbox{} &=& \lim_{u'\rightarrow u}~\frac{1}{(u+u')^2} =
 \frac{1}{4u^2} \ .
\label{comm} \eea
However, the
result is divergent on the null line $u=0$. The problem arises
from the non-trivial geodesics
which can also become zero length (see Fig. \ref{fig:figgg5.ps}).
A similar calculation
for the left-movers yields a divergence at $v=0$. Hence one
concludes that the orbifold is potentially unstable. Similar
calculations can be done in higher dimensions.

However, upon closer inspection the above argument has some
puzzling features. If we want to associate the two-point function
(\ref{gorb}) with a field operator, the operator should be
symmetric under the $u\rightarrow -u$ $\Zmath_2$ reflection.
A naive way to impose the invariance is to consider
\begin{equation}
  \tilde{\phi}(u) = \frac{1}{\sqrt{2}} (\phi (u)+\phi (-u)) \ .
\label{phitilde}
\end{equation}
Formally, one can check that the renormalized expectation value
(\ref{comm}) is that of the $\Zmath_2$ invariant
field operator, with the four contributions associated
with 'short' and 'long' contractions.
\myfig{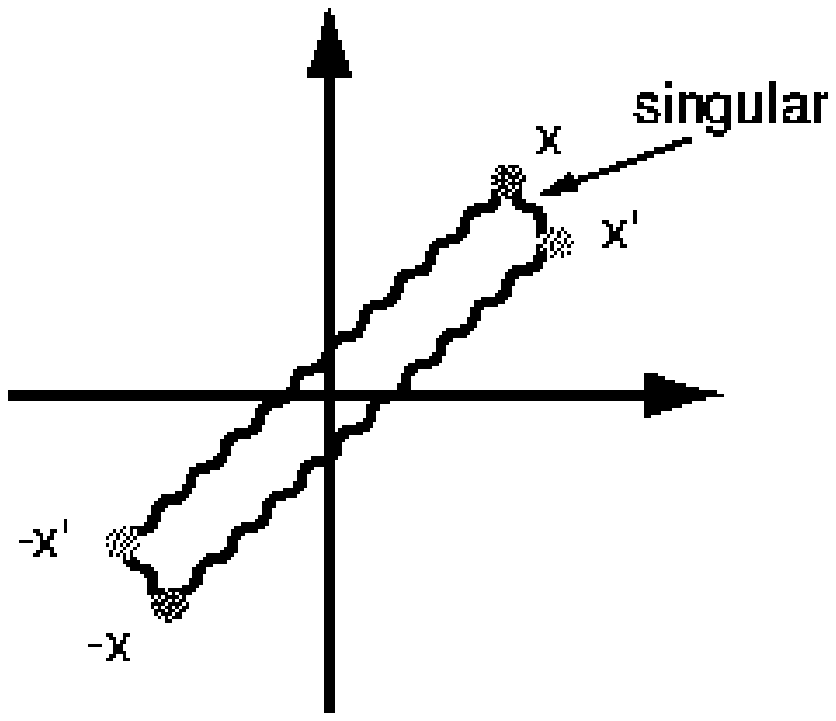}{4}{Correlator of point-split
composite operator. The 'short' contractions, between $x$ and $x'$ are
the usual
short-distance ones, and should be subtracted. The 'long' contractions
give rise to the Casimir energy.}
However, this construction has various problems.
The most cumbersome one is that the $\Zmath_2$
invariant field operator (\ref{phitilde}) has the mode expansion
\begin{equation}
 \tilde{\phi}(u) = \sqrt{2}
 \int d\omega~(a_\omega + a^\dagger_\omega)\cos(\omega u)
\end{equation}
so it is not clear what exactly is meant by the naive notion
of particles and vacuum. The problem of constructing a global
Fock space is also well known from investigations of
elliptic de Sitter space $dS/\Zmath_2$
\cite{Schrodinger1,Gibbons:1986dd,Folacci:1987gr,Parikh:2002py}.
In the above, the problem has been lifted onto $\orbid$, where
the $dS/\Zmath_2$ can be embedded.

Actually, we will argue that the orbifold
identification requires identifying a particle with positive energy
at $(t,x)$ with a particle with negative energy at $(-t,-x)$. Particles
of the latter kind cannot be created with $a^\dagger_\omega$. A quick
look at the mode expansion of $\phi (-u)$ might give a false impression
that this would happen, but really $\phi (-u)$ is just the field operator
$\phi$ evaluated at point $-u$ rather than a new operator with the
creation and annihilation operators acting in a different way.
Another problem is that the usual
prescription calls us to evaluate commutators of field operators
at equal time. On the orbifold covering space this becomes
problematic, since ``equal time" now corresponds to times $t$
and $-t$. For these reasons we would like to take a step
back and reconsider the formulation of field theory on the
$\Rmath^{1,d}/\Zmath_2$ orbifold.
However, we will first examine if the divergence of the
stress tensor persists in string theory. The result that we find
will provide additional motivation to reconsider the formulation
of field theory.

\section{The String Theory Calculation}
\label{sec:stringcalc}

Our next goal is to calculate the backreaction on the orbifold
at one-loop level in string theory. In practice, this is done by
calculating the one-loop graviton tadpole.

If we write the metric tensor as
$g_{\mu\nu}(x)= \eta_{\mu\nu}+2\kappa h_{\mu\nu}(x)$, the
vev of the stress tensor may be written \cite{Birrell:1982ix}
\begin{equation}
\bra T_{\mu\nu}\ket = -i\frac{\delta}{\delta g^{\mu\nu}}\ln
Z^{2nd}_{EFT}|_{h^{\mu\nu}=0} = -\frac{i}{2\kappa}\frac{\delta
Z_{1st}}{\delta h^{\mu\nu}}|_{h^{\mu\nu}=0} \ . \label{T1}
\end{equation}
In the above, we used the relation between the vacuum amplitudes
in the second quantized and first quantized formalism, $Z_{2nd} =
e^{Z_{1st}}$, to replace the effective field theory action $\ln
Z^{2nd}_{EFT}$ by the point particle partition function $Z_{1st}$.

Now we replace point particles by strings. At one-loop level
\cite{bigbook1}
\begin{equation}
  Z^{ST}_{1-loop}[g] = \int \frac{d\tau d\bar{\tau}}{4\tau_2}~Z(\tau)
  = \int \frac{d\tau d\bar{\tau}}{4\tau_2} \int_{T^2}
  \Dcal X~e^{i\frac{T}{2}\int d^2w~g_{\mu\nu}(X)\pat X^\nu \patb
  X^\nu} \ .
\end{equation}
This is then inserted in (\ref{T1}). \footnote{This is somewhat
reminiscent of a recent calculation in \cite{Larsen:2002wc}.}
Suppressing the integral over $\tau$, we have
\begin{equation}
  Z^{ST}_{1-loop} = \int
  \Dcal X~e^{i\frac{T}{2}\int d^2w~\eta_{\mu\nu}\pat X^\mu \patb X^\nu}
  \left\{ 1+i\frac{g_{str}}{\alpha'}\int d^2w~h_{\mu\nu}(X)\pat X^\mu \patb
  X^\nu +\cdots \right\} \ .
\end{equation}

Now Fourier expand the perturbation,
\begin{equation}
 h_{\mu\nu}(X) = \int \frac{d^{D+1}k}{(2\pi)^{D+1}}~e_{\mu\nu}(k)e^{ik\cdot X}
\end{equation}
and introduce
\begin{equation}
  V_{\mu\nu}(k) = \pat X^\mu \patb X^\nu e^{ik\cdot X} \ ,
\end{equation}
then
\begin{equation}
  Z^{ST}_{1-loop} [g] =
   Z^{ST}_{1-loop} [\eta ] +
  i\frac{g_{str}}{\alpha'} \int \frac{d^{26}k}{(2\pi)^{D+1}}\int d^2w~e_{\mu\nu}(k)
  \bra V^{\mu\nu}(k;w) \ket +\cdots \ .
\label{Z}
\end{equation}
We then get
\begin{equation}
\bra T^{\mu\nu}(x) \ket = \frac{1}{4\pi \alpha'}
\int \frac{d^{D+1}k}{(2\pi)^{D+1}}\int d^2w
\bra V^{\mu\nu}(k;w) \ket e^{-ik\cdot x}
\end{equation}
the relation between the Fourier transformed tadpole and the
stress tensor.

Note that in Minkowski space, one obtains
\begin{equation}
 \bra V^{\mu\nu}(k)\ket = -\left( \frac{g_{str}}{4\pi\tau_2} \right)
 \frac{\delta^{(D+1)}(\sqrt{\alpha'}k)}{V_{D+1}}
 \eta^{\mu\nu} Z_{1-loop} \ ,
\end{equation}
so that
\begin{equation}
  \bra T_{\mu\nu}\ket \sim \frac{1}{{\alpha'}^{13}V_{26}} Z_{1-loop} \times
  \eta_{\mu\nu}
\end{equation}
which is of the right form for the stress tensor of a cosmological
constant $\Lambda \sim Z_{1-loop}$.

On $\Zmath_2$ orbifolds, the story is essentially similar.
What is different in the string graviton tadpole calculation is
that a) the relevant vertex operator must be $\Zmath_2$ invariant: it
is the sum of vertex operators carrying $k$ and $-k$ in the
directions of the orbifold, and b) a priori there are contributions
from the twisted sector strings. The fact a) suggests that the Fourier
transform of the tadpole will be the sum
\begin{equation}
   \bra T_{\mu\nu}(X)\ket + \bra T_{\mu\nu}(-X)\ket ,
\label{Tsum}
\end{equation}
where $X$ are the coordinates along the orbifold directions.
This could be obtained from the effective action by including the functional
differentiation $\delta /\delta h_{\mu\nu}(-X)$. We will return to
these issues when we discuss the reformulation of QFT. Let us
first proceed with the calculation of the tadpole.

\subsection{One-loop Graviton Tadpole}

Now we proceed to give some of the details of the calculation of the
one-loop graviton tadpole in string theory described above.
Our calculations are based on the functional method.
We begin with a brief review of the latter, following \cite{bigbook1}.
As it turns out, an immediate difference with tadpole calculations
on Euclidean orbifolds is in kinematics and in appropriate choice of polarization
of vertex operators.
We have also performed
the same calculations in the oscillator formalism. It also turns out
that there are some interesting subtleties and differences with the
standard discussion; detailed notes may be found in
the appendix.

We should note that in the string computations, one usually
performs a Wick rotation in both spacetime and worldsheet,
necessary for formal convergence.
If the target space is time-dependent,
the standard techniques of analytic continuation may not be
applicable.\footnote{See {\em e.g.} \cite{Mathur:1993tp} for a proposal
to modify the standard approach.}
In the context of the $\orbid$ orbifold, the
issue was already noted in \cite{vijayetal}. In the present paper,
we simply adopt the same strategy as in \cite{vijayetal}, namely
we formally continue the worldsheet to Euclidean signature in the
calculations to obtain
an expression for the tadpole. As well, we will encounter zero-mode
integrations whose values are defined by a spacetime Euclidean
continuation. The result is apparently well-defined and in a later
section, we search for a field theory formalism that is compatible
with the low-energy limit. In that section, propagation on the
orbifold will be essentially shown to be an identification of forward and
backward propagation on the covering space $\Rmath^{1,d}$. This
may also explain why the formal analytic continuation prescription
continues to work in the calculations of this section.

\subsection{The Generating Functional on $\Rmath^{1,d-1}$}

Following \cite{bigbook1}, the generating functional is
\begin{equation}
  Z[J] = \bra \exp \{ i\int d^2w J_\mu (w,\wbar ) X^{\mu}(w,\wbar )
  \} \ket \ .
\label{genfun}
\end{equation}
In order to perform the functional integrals, we introduce a
complete set eigenmodes $X_I$ of the Laplacian $\nabla^2$ on the
toroidal worldsheet,
\bea
  && \nabla_w^2 X_I (w,\wbar ) = -\omega^2_I X_I (w,\wbar )\ \ \ \  , \nonumber \\
 &&  \int d^2w~X_I(w,\wbar)X_J(w,\wbar )= \delta_{IJ} \
\eea
and expand the string embedding coordinates in the eigenmodes,
\begin{equation}\label{eq:modes}
  X^\mu (w,\wbar ) = \sqrt{4\pi^2\alpha'}\sum_I x^\mu_I X_I(w,\wbar )\ .
\end{equation}
We also denote
\begin{equation}
      J_{\mu, I}=  \sqrt{4\pi^2\alpha'}\int d^2 w J_\mu (w,\wbar ) X_I(w,\wbar ) \ .
\end{equation}
We then integrate out the expansion coefficients $x^\mu_I$ by completing
the squares in the generating
functional and performing the resulting Gaussian integrals. In particular,
the integrals will include zero mode contributions from
$x^\mu_0$. The result in $d$ target space dimensions is
\begin{equation}
  Z[J]= N[J_0]\left[ {\det}'  \left( -\nabla_w^2\right) \right]^{-d/2}
  \exp \left\{ -\half \int d^2 w
  \int d^2 w' J(w)\cdot G'(w,w')\cdot J(w') \right\} \ ,
\end{equation}
where $N[J_0]$ is the zero mode contribution
\begin{equation}
  N[J_0]= i(2\pi)^d \delta^{(d)}(J_0) \ ,
\end{equation}
(with $i$ coming from the Wick rotation $x^0_I\equiv ix^d_I$), the
determinant factor is
\begin{equation}
  {\det}'   \left( -\nabla_w^2\right) \equiv \prod_{I\neq 0} \omega^2_I\ ,
\label{det}
\end{equation}
and $G'(w,w')$ is the Green function
\begin{equation}
   G'(w,w') = \sum_{I\neq 0} \frac{2\pi\alpha'}{\omega^2_I}X_I(w)X_I(w') \ .
\end{equation}
The latter satisfies the differential equation
\begin{equation}
   -\frac{1}{2\pi\alpha'}\nabla^2_w G'(w,w') = g^{-1/2}\delta^{(2)}(w-w')-X^2_0 \ ,
\label{gfeqn}
\end{equation}
where $X_0$ is the zero mode of the Laplacian on the torus. The
functional determinant (\ref{det}) gives the torus partition function,
\begin{equation}
   Z_{T^2} [0] = V_d\left[ \alpha' X_0^2{\det}'
   \left(-\nabla_w^2\right) \right]^{-d/2}
\end{equation}

\subsection{The Generating Functional on Orbifolds}

Next we generalize this to the case of the orbifold. For
comparison, we will consider two related types of orbifolds:
\begin{description}
\item [A)] The Euclidean orbifold $\mathbb{R}^{1,d}\times \mathbb{R}^{25-d}/
\mathbb{Z}_2$
\item [B)] The Lorentzian orbifold $\mathbb{R}^{1,d}/\Zmath_2 \times
\mathbb{R}^{25-d}$.
\end{description}
To streamline the notation, we will denote the total number of
orbifold directions in both cases as $d_o$.
We split the coordinates $X$ and the components of the
source $J$ into those along the orbifolded ($o$) and un-orbifolded ($u$)
directions.
The generating functional takes the form
\begin{equation}
  Z[J] = \sum^1_{g=0}\sum^1_{h=0}~\langle \exp \{ i\int J_{o}
  \cdot X_{o}
     + i\int J_u \cdot X_u \} \rangle_{gh}
\label{genfuno}
\end{equation}
including the sum over the untwisted ($g=1$) and twisted ($g=0$)
sectors, with ($h=0$) and without ($h=1$) the $\Zmath_2$
reflection, for string oscillations in the orbifolded directions.
We then again expand $X^\mu$ in the eigenmodes of $\nabla^2$, but
now the eigenvalues and -modes will be different in the orbifolded
directions for each sector, due to the different (anti)periodic boundary
conditions. After integrating over the eigenmode coefficients, the
functional takes the form
\bea\label{genfunoans}
  Z[J] &=& \frac{N_u [J_0]}{N_u [0]} Z_u[0]~\exp\{-\half
  \int d^2w\int d^2w'
  J_u(w)\cdot J_u(w')~G'(w,w')\}  \\
  \mbox{} &\times&
  \sum_{gh}~\frac{N_{o,gh}[J_0]}{N_{o,gh}[0]}Z_{o,(g,h)}[0]~\exp
  \{-\half \int d^2w\int d^2w'
  J_{o}(w)\cdot J_{o}(w')~G'_{(g,h)}(w,w')\} .\nonumber
\eea
In the above, $N_u[J_0]$, $N_{o,(g,h)}[J_0]$ are the zero mode
contributions (we have formally multiplied and divided by $N[0]$ recognizing that $Z$ includes such a factor.)
In the orbifolded directions, there are zero modes only in the untwisted sector
without the $\Zmath_2$ reflection, and none in the other sectors
because $X$ satisfies an antiperiodic boundary condition in at
least one of the toroidal worldsheet directions. Thus, for $J=k\delta^{(2)}
(w-w')$,
\begin{equation}
  \frac{N_{o,(1,1)}[J_0]}{N_{o,(1,1)}[0]}
  =\frac{1}{V_{d_o}}\delta^{(d_o)}(k) \ ;
  \ N_{o,(g,h)}[k] = 1\ {\rm for}\
  (g,h)\neq (1,1) \ .
\end{equation}
The factors $Z_u[0]$, $Z_{o,(g,h)}$ are the partition
function contributions
from the directions transverse to and parallel
with the orbifold, including the four untwisted and twisted
$(g,h)$-sectors. Explicitly \cite{vijayetal},
\bea
  Z_{o,(1,1)} &=& \frac{V_{d_o}}{2}~\left|
  \frac{1}{\sqrt{\tau_2}~\eta^2 (\tau)}
  \right|^{d_o} \nonumber \\
  Z_{o,(g,h)} &=& \left| \frac{\eta (\tau )}{\theta_{gh} (\tau )}
  \right|^{d_o} ,\ \ \ \ \ (g,h)\neq (1,1)
\eea
There are four different Green functions, corresponding to the
different periodicities on the toroidal worldsheet. The doubly
periodic one is \cite{bigbook1}
\begin{equation}
 G'_{(1,1)}(w,w')\equiv G'(w,w')=
  -\frac{\alpha'}{2}\ln \left| \theta_{11}\left(\left.\frac{w-w'}{2\pi}
  \right|\tau \right)
  \right|^2 + \pi\alpha' X_0^2 [{\rm Im}(w-w')]^2\ ,
\end{equation}
and the other ones with at least one antiperiodic direction are
\bea
  G'_{(1,0)}(w,w') &=& -\frac{\alpha'}{2} \ln \left|
  \frac{\theta_{11} (\frac{w-w'}{4\pi }|\tau )\theta_{10}(\frac{w-w'}{4\pi}|\tau )}
  {\theta_{00} (\frac{w-w'}{4\pi }|\tau )\theta_{01}(\frac{w-w'}{4\pi}|\tau )}
  \right|^2
  \nonumber \\
  G'_{(0,1)}(w,w') &=& -\frac{\alpha'}{2} \ln \left|
  \frac{\theta_{11} (\frac{w-w'}{4\pi }|\tau )\theta_{01}(\frac{w-w'}{4\pi}|\tau )}
  {\theta_{10} (\frac{w-w'}{4\pi }|\tau )\theta_{00}(\frac{w-w'}{4\pi}|\tau )}
  \right|^2 \nonumber \\
  G'_{(0,0)}(w,w') &=& -\frac{\alpha'}{2} \ln \left|
  \frac{\theta_{11} (\frac{w-w'}{4\pi }|\tau )\theta_{00}(\frac{w-w'}{4\pi}|\tau )}
  {\theta_{01}(\frac{w-w'}{4\pi}|\tau )\theta_{10} (\frac{w-w'}{4\pi }|\tau )}
  \right|^2 \ .
\eea
In $n$-point amplitudes, one also encounters
self-contractions which require renormalization. A simple
prescription is to subtract the divergent part $-\frac{\alpha'}{2}
\ln|w-w'|^2$ from the Green functions and define their renormalized
versions. The renormalized version of $G'_{11}$ is \cite{bigbook1}
\begin{equation}
  G'_{(1,1),ren}(w,w)= -\frac{\alpha'}{2}\ln \left| \frac{\theta'_1(0|\tau )}
  {2\pi} \right|^2 \ .
\label{g11ren}
\end{equation}
After some manipulations, the renormalized versions of the other
Green functions also turn out to simplify considerably to the
following simple forms:
\begin{equation}
  G'_{(g,h),ren} =-\frac{\alpha'}{2}\ln |\theta_{gh}(0|\tau)|^4
\end{equation}
for $(g,h)\neq (1,1)$.

\subsection{One-loop Graviton Tadpole on the Orbifold}

Consider then the one-loop graviton tadpole
on the orbifold. The vertex operator for a state which is not
projected out by the $\Zmath_2$ reflection must be symmetric under $X\rightarrow
-X$, hence the relevant massless tadpole on the orbifold is
\bea
&& \bra V_{\mu\nu}(k_{o},k_u)
+ V_{\mu\nu}(-k_{o},k_u) \ket  \nonumber \\
&& \ \ \ \ \ \ \ \ \ \ \mbox{}=   \frac{2g_{str}}{\alpha'}
\bra \pat X^\mu \patb X^\nu e^{ik_{o}\cdot \Xpar +ik_u \cdot \Xperp }
+ \pat X^\mu \patb X^\nu e^{-ik_{o}\cdot \Xpar +ik_u \cdot \Xperp }
      \ket
\eea
The momentum must satisfy the on-shell condition
$k^2=-m^2=0$. Now there are some immediate choices to be done where
the Euclidean and Lorentzian
orbifolds {\bf A} and {\bf B} differ. In string theory one often
considers Euclidean orbifolds as a way of compactifying extra
dimensions. Therefore one is usually interested in states which only
propagate and carry polarization in the non-orbifolded noncompact
directions, and
the momentum and the polarization are chosen to be entirely transverse to the
orbifold, with $k^2=k^2_u=-m^2$. However, in the Lorentzian
orbifold one must also include momentum components in orbifold directions
in order to satisfy the on-shell condition.
Furthermore, in the Lorentzian case, in order to compare with
the quantum field theory calculation of Section
\ref{sec:tradcalc}, we choose the polarization to be along the
orbifold directions\footnote{Another reason why this is the interesting
case is to view the orbifold as a cosmological toy model. If one would make
the model truly $d+1$-dimensional, the extra dimensions would need to be
compactified. The massless gravitons would carry polarization in the
non-compact orbifold directions.}.

We evaluate the tadpole by first performing a point splitting and then
functional differentiation of the generating functional,
\bea
  && \bra \pat X^\mu (w,\wbar )\patb X^\nu (w,\wbar ) e^{ikX(w,\wbar)} \ket
  = \nonumber \\
 &&  \ \ \ \ \ \ (-i)^2\lim_{w_1,w_2\rightarrow w} \pat_{w_1}
 \patb_{w_2}~\frac{\delta}{\delta J_\mu (w_1)}
 \frac{\delta}{\delta J_\nu (w_2)}~\bra
 \exp \{i\int d^2 w' J_\lambda (w')X^\lambda (w')\} \ket \ ,
\eea
evaluated at $J (w')=k~\delta^{(2)} (w'-w)$.
Before the functional differentiation, for
the generating functional we substitute the integrated form
(\ref{genfunoans}). We will also substitute the on-shell condition
$k^2=0$.

In the case where the polarizations are in the unorbifolded directions,
the functional differentiation and the on-shell condition gives
\bea
  && \bra \pat X^\mu (w) \patb X^\nu (w) e^{ikX} \ket = \nonumber
  \\
  && \ \ \frac{N_u[k]}{N_u[0]}Z_u[\tau]~\lim_{w_1,w_2\rightarrow w}
  \left[~\eta^{\mu\nu}\pat_{w_1}\patb_{w_2}
  G'(w_1,w_2)-k^\mu k^\nu~\pat_{w_1}G'(w_1,w)\patb_{w_2}G'(w_1,w)\right]
  \nonumber \\
  && \ \ \times  \sum_{g,h} Z_{o,(g,h)}[\tau ]
\label{euc1}
\eea
whereas when the polarizations are in orbifolded directions, the corresponding result is
\bea
 && \bra \pat X^\mu (w) \patb X^\nu (w) e^{ikX} \ket = \nonumber
  \\
  && \ \ \frac{N_u[k]}{N_u[0]}Z_u[\tau]
  \times \sum_{g,h} \frac{N_{o,(g,h)}[k]}{N_{o,(g,h)}[0]}Z_{o,(g,h)}[\tau ] \nonumber \\
 &&\ \ \ \ \ \lim_{w_1,w_2\rightarrow w} \left[~\eta^{\mu\nu}\pat_{w_1}\patb_{w_2}
  G'_{(g,h)}(w_1,w_2)-k^\mu k^\nu~\pat_{w_1}G'_{(g,h)}(w_1,w)\patb_{w_2}G'_{(g,h)}(w_1,w)\right]
  \ .
\label{lor1}
\eea
In both cases, the Green function will need to be replaced by
their renormalized versions.
We can already see that the expressions are quite different. Let
us simplify them further. First, we can use the equation
(\ref{gfeqn}) to simplify the double derivatives of the Green
functions. First, since $G'(w_1,w_2)=G'(w_1-w_2)$,
\begin{equation}
  \pat_{w_1}\patb_{w_2} G'(w_1,w_2)=
  -\pat_{w_1}\patb_{w_1}G'(w_1,w_2)\ .
\end{equation}
On the other hand, the equation (\ref{gfeqn}) evaluates to
\begin{equation}
 \pat_w\patb_w G'(w,w') =
 -\pi\alpha'\delta^2(w-w')+\frac{\pi\alpha'}{2}X^2_0 \ .
\end{equation}
The first term on the right hand side
originates from the short distance divergence $G'(w_1,w_2)\sim \ln|w_1-w_2|^2$
of the Green function, which we subtract off when we renormalize
the Green functions. The latter then satisfy the equation
\begin{equation}
  \pat_{w_1}\patb_{w_2}G'_{ren}(w_1,w_2) = -\frac{\pi\alpha'}{2}X^2_0 \ .
\end{equation}
Similar results hold for the renormalized Green functions $G'_{(g,h),ren}$. Since a zero mode $X_0$ exists
only in the doubly periodic $(g,h)=(1,1)$ sector, the double derivatives
$\pat\patb G'_{gh,ren}$ vanish in all the other three sectors.

Next, we examine the first derivatives of the renormalized Green
functions. A short calculation shows that in all cases the Green functions
have a short distance behavior of the type
\begin{equation}
   \pat G'_{(g,h)} (w,w') \approx_{w\rightarrow w'} -\frac{\alpha'}{2}(w-w')^{-1}
   + C_{(g,h)}(\tau) (w-w') + {\cal O}((w-w')^3)
\end{equation}
where $C_{(g,h)}(\tau)$ are rational functions of
derivatives of theta functions at $(0|\tau)$. A similar formula is
found for the antiholomorphic derivative $\patb G'_{(g,h)}$.
There is only one divergent term, due to the self-contraction
of $X$ with $\pat X$. The renormalization prescription again removes the
divergent term, so the renormalized (derivatives of) Green function
vanish in the limit $w\rightarrow w'$. Hence these terms will not
contribute to the graviton tadpole.

Substituting all the normalization and partition function factors,
the final results are
\begin{equation}\label{tadpu}
  \bra V^{\mu \nu}(k)+  V^{\mu \nu}(-k) \ket_{1-loop} =
 -\frac{g_{str}}{4\pi\tau_2} \frac{\delta^{(d_u)}(k)}{V_u}
 Z_u[\tau]\times\sum_{g,h} \frac{N_{o,(g,h)}[k]}{N_{o,(g,h)}[0]}Z_{o,(g,h)}[\tau ]~\eta^{\mu\nu} \ .
\end{equation}
for polarizations in the unorbifolded directions, and
\begin{equation}\label{tadpo}
  \bra V^{\mu \nu}(k)+  V^{\mu \nu}(-k) \ket_{1-loop} =
 -\frac{g_{str}}{4\pi\tau_2}
 \frac{\delta^{(26)}(k)}{V_{26}}~Z_u(\tau)
 Z_{o, (1,1)}[\tau]~\eta^{\mu\nu} \ .
\end{equation}
for polarizations in the orbifolded directions. By
analogy, one would then expect this tadpole to vanish for the
superstring.

Equation (\ref{tadpu}) is the standard result.
In the case of the Lorentzian orbifold, we would like to think of
spacetime as the orbifolded directions, while the unorbifolded directions
are perhaps compactified. Thus in the Lorentzian orbifold, it is appropriate
to consider (\ref{tadpo}).
At first sight, this result looks rather surprising, as it is precisely the
same as for a graviton in the usual $\Rmath^{1,25}$ target space.

This is in direct conflict with the field theory calculation of the previous
section, but we have already noted the problems of principal with that
calculation. In the light of the string theory analysis, we must search for a
field theory description that can be consistent with these
results. A key observation is that the string calculation involved
strings with $k$ and $-k$, opposite spacelike momentum {\em and}
energy.

\section{Quantum Field Theory on $\orbid$ Revisited}
\label{sec:qftnew}

Consider a point particle on the fundamental domain of $\orbid$.
On the covering space, it corresponds to two particles: one with
positive energy (propagating forward in time), and its image
with negative energy (propagating backward in time) with opposite
momentum (Fig. \ref{fig:partic.eps}). In other words, for each particle with a
momentum $(k^0,\vek)$ we must include its image with momentum
$(-k^0,-\vek)$.

\myfig{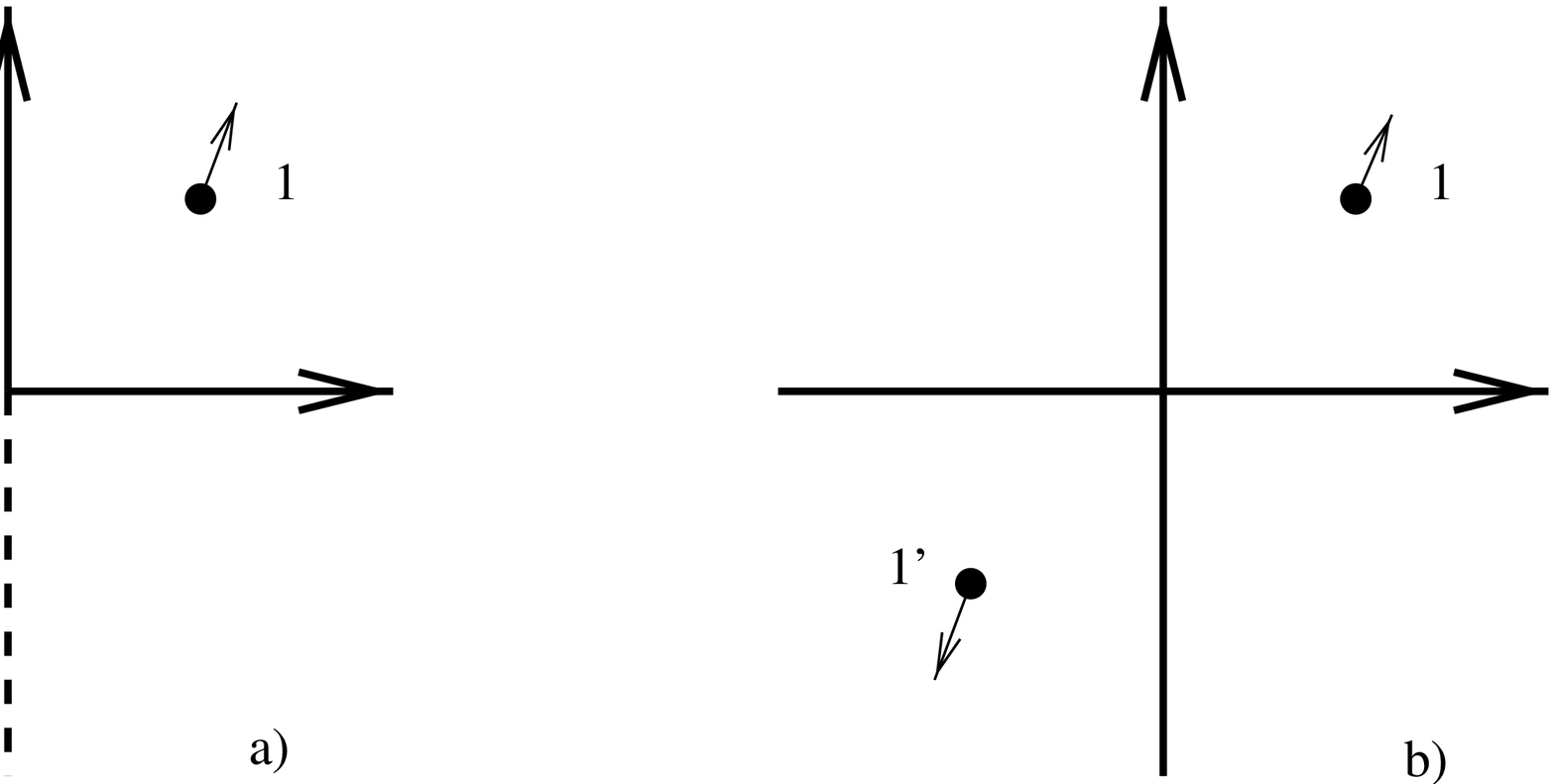}{7}{A point particle on the orbifold.
a) depicts a single point particle on the fundamental domain,
while b) depicts the point particle and its image on the
covering space, moving towards the opposite
time and space directions.}

The situation is similar for strings on $\orbid$, as analyzed in
\cite{vijayetal}. The
states in the untwisted sector which survive the $\Zmath_2$
projection are of the type
\begin{equation}
  |\psi \ket_S = \left( \alpha^{\mu_1}_{-n_1}\cdots
  \tilde{\alpha}^{\nu_1}_{-m_1}\cdots \right)_{S,A}~(|0,k\ket
  \pm |0,-k\ket ) \ ,
\end{equation}
{\em i.e.} symmetrized combinations of string states with
opposite pairs of center-of-mass momentum $k$.
We conclude that in order to have a good description of quantum
mechanics on the orbifold, we must start with pairs of states
with opposite energy and momentum.

In quantum field theory in Minkowski space,
1-particle states in the Fock space are associated with
positive energy,
\begin{equation}
   |\omega_k , \vek \ket = a^{\dagger}_{\vek} | 0 \ket \ \ {\rm
   with} \ k^0 = \omega_k >0.
\end{equation}
We should think of this as a projection of the Hilbert space containing both positive and negative energies.
In a Euclidean orbifold, there would be an additional projection onto invariant states.

However, in order to formulate
a quantum field theory on the Lorentzian orbifold covering space,
we must also be able to include states with
$k^0 =-\omega_k <0$. In other words, what we need is a Fock space
$\Hcal$ which involves sectors with both sign choices for the
energy:
\bea
  \Hcal^+ &=& \prod_{\vek} \Hcal^+_{\vek} \ \ {\rm with} \ k^0
  =\omega_k > 0 \nonumber \\
  \Hcal^- &=& \prod_{\vek} \Hcal^-_{\vek} \ \ {\rm with} \ k^0
  =-\omega_k < 0 \ .
\eea
The full Fock space is then the direct sum
\begin{equation}
  \Hcal = \Hcal^+ \oplus \Hcal^- \ .
\end{equation}
This is an essential difference with usual Euclidean
orbifolds. What survives on the orbifold QFT is the invariant Fock
space. In the Euclidean case we can start directly with the usual
Fock space $\Hcal^+$ and project out the non-invariant states.
However, in our case, in order to construct an invariant Fock
space, we first need to extend the Fock space to include the $\Hcal^-$
sector. Next, to find the invariant states on the orbifold, we
need to first implement the
$\Zmath_2$ action as an isomorphism $\Hcal^\pm \rightarrow
\Hcal^\mp$ which acts by flipping the sign of energy
and momentum in the orbifolded directions. In particular, the
usual vacuum $|0\ket\in \Hcal^+$ must map to a state in $\Hcal^-$; we
will call it $|\tilde{0}\ket$. We will later define it and other
states in $\Hcal^-$ more
precisely. The invariant Fock space is then
\begin{equation}
    \Hcal_{inv} = \Hcal / \Zmath_2 \ .
\end{equation}
Given this orbifold identification, it should be noted that there is
no particular problem with the stability of the theory related to a negative
energy sea.


Let us introduce two sets of annihilation and creation operators which at
least provisionally commute with one another
\begin{equation}
   [a_{\vek},\atilde_{\vek'}]=[a_{\vek},\atilde^\dagger_{\vek'}]
   = [a^{\dagger}_{\vek},\atilde_{\vek'}]
   =[a^{\dagger}_{\vek},\atilde^\dagger_{\vek'}]=0 \ .
\end{equation}
with
\begin{equation}
  [a_{\vek},a^{\dagger}_{\vek'}]=  [\atilde_{\vek},\atilde^{\dagger}_{\vek'}] =
  (2\pi)^d\delta^{(d)}(\vek -\vek') \ .
\end{equation}
We envision that  $a_{\vek}$ destroys a particle with wavefunction
$e^{-i\omega_{\vek} t+i\vek
\cdot \vex}$ which is positive energy and momentum with respect to the Killing
vectors $E=i\pat_t$ and $P=-i\nabla$, whereas
$\atilde_{\vek}$ destroys a particle with wavefunction $e^{i\omega_{\vek}t
-i\vek \cdot \vex}$ which is negative energy and opposite momentum
with respect to $E$ and $P$. Let us then define a new vacuum $|\tilde{0}\ket$
and 1-particle states:
\bea
 |\tilde{0}\ket\ &:& \ \atilde_{\vek}|\tilde{0}\ket =0 \nonumber
 \\
 |-\omega_{\vek},-\vek\ket \ &:& \
|-\omega_{\vek},-\vek\ket = \atilde^{\dagger}_{\vek}|\tilde{0}\ket
\ .
\eea
We use the notation $|-\omega_{\vek},-\vek\ket$ to
emphasize that these particles carry negative energy and opposite
momentum. The above states are the images of the usual vacuum and
1-particle states of $\Hcal^+$ under the $\Zmath_2$ isomorphism
$\Hcal^+\rightarrow \Hcal^-$, that we discussed earlier.

Next we need to take into account the identification and define
$\Zmath_2$ invariant states on the fundamental domain; these are
states in the invariant Fock space $\Hcal_{inv}=(\Hcal^+\oplus
\Hcal^-)/\Zmath_2$. {\em E.g.} for 1-particle states we define
\begin{equation}
     |\omega_{\vek},\vek\ket_{inv}
      = \frac{1}{\sqrt{2}}
     \left( \begin{array}{l} |+\omega_{\vek},+\vek\ket \\
     |-\omegak ,-\vek\ket \end{array} \right) \ .
\end{equation} Invariant multiparticle states are constructed in an analogous
fashion.

We propose that the natural energy operator on the orbifold is
\begin{equation}
 H_{inv} = \sum_{\vek}N_{\vek}\omega_{\vek}
 \mbox{} = \left( \begin{array}{cc}
 \sum_{\vek} \omega_{\vek} a^\dagger_{\vek}a_{\vek} & \mbox{} \\
 \mbox{} & \sum_{\vek} \omega_{\vek}\atilde^\dagger_{\vek}
 \atilde_{\vek}  \end{array} \right)
 = \left( \begin{array}{cc} H^+ & \mbox{} \\
\mbox{} & H^- \end{array} \right) \ .
\label{enop}
 \end{equation}
The individual pieces $H^\pm$ generate time
translations in $\pm t$ directions. The orbifold identifies the
two, hence on the covering space neither direction is preferred. So
the theory on the covering space must start with a symmetric
combination of the two Hamiltonians $H^\pm$.
Including also the zero point energies in $H^{\pm}$
(the $\Zmath_2$ invariance
also extends to the zero energy contributions), we can then evaluate
the vacuum energy on the orbifold,
\bea
  {}_{inv}\bra 0 | H_{inv} |0\ket_{inv} &=& \half \bra 0 | H^+ |0\ket
  + \half \bra \tilde{0} | H^- | \tilde{0} \ket
  \nonumber \\
  \mbox{} &=& \half \bra 0 | \half \sum_{\vek}\omegak
  |0\ket + \half \bra \tilde{0} | \half \sum_{\vek}\omegak
  | \tilde{0} \ket = \half \sum_{\vek}\omegak \ .
\eea
This is the usual vacuum divergence.

The invariant Hamiltonian operator
(\ref{enop})
must derive from an invariant stress tensor. Thinking
of the latter as an operator, it will also reduce to components
which act on the subspaces $\Hcal^\pm$. In our notation, the
invariant stress tensor should be written
\begin{equation}
  T^{inv}_{\mu\nu}(t,\vex) = \left( \begin{array}{cc}
  T^+_{\mu\nu} (t,\vex) & \mbox{} \\ \mbox{} & T^-_{\mu\nu} (-t,-\vex)
  \end{array} \right)
\mbox{}
\end{equation}
We would like to give a field theoretic description of such a calculation.
Given the structure of the Fock space, it appears natural to describe
the system using a pair of scalar fields which at least in some approximation
do not interact with one another. Such doubling of the degrees of freedom
seems to be a common occurrence in time-dependent backgrounds \cite{Mathur:1993tp}.
The novelty here is not so much this doubling, but the fact that we must deal
carefully with the orbifold identification.

Finally, we comment on the number of degrees of freedom. Let us compare with
$\Rmath \times (\Rmath^d / \Zmath_2)$ where $\Rmath$
is the time direction. In
that case, the invariant Fock space has half as many degrees
of freedom as the full Fock space. In $\orbid$, we first doubled
the degrees of freedom and then projected out half of them, so the
remaining number of states in $(\Hcal^+ \oplus \Hcal^-)/\Zmath_2$ is
the same as in Minkowski space.
However, recall the discussion after Fig. 2, on the freedom to choose the
time direction on the fundamental domain. In order for the states
not to propagate through the big crunch and continue on to the reversed
time orientation, an additional projection would be needed. That
projection would presumably again project out half of the states, so the
remaining number would be in agreement with that in the Euclidean
$\Rmath \times (\Rmath^d / \Zmath_2)$ orbifold.
Essentially the projection should correspond to
some sort of a boundary condition, presumably near the initial
value surface. We don't know how to implement this precisely, but we
will make some additional comments on this in the next section and
in Section \ref{sec:smat}.


\subsection{A Field Realization}
There are of course other well known reasons to involve
both positive and negative energy sectors in the formulation of
QFT. One of them is QFT in curved spacetime, or other cases
where we compare observers who are not related by proper orthochronous
Lorentz
transformations.  The mode expansions of the field operator
relevant for such observers are related by mixing of positive
and negative energies. In the present case, the $\orbid$ spacetime
is locally flat, but we want to identify (as opposed to compare) observers
related by the time (and space) reflection, and identify
the corresponding degrees of freedom.

Actually, a closely related starting point is QFT in flat
spacetime but at finite temperature (FTQFT). The real-time
formulation of FTQFT also leads to mixing between positive and
negative energies. For inspiration, we shall review it briefly.
The starting point in the path integral formulation of real-time
FTQFT is the generating functional
\begin{equation}
  Z[J] = \int \Dcal \phi~\exp \left\{ i \int_C d^4x~[\Lcal (\phi (x))
  + J(x)\phi (x)] \right\}
\label{ZFT}
\end{equation}
where the time integral has been promoted to a contour integral
along a complex time path $C$, starting from some initial time
$t_i$ and ending at a complex final time $t_i-i\beta$, where the
imaginary part is given by the inverse temperature $\beta =T^{-1}$
\cite{Niemi:1984nf}. The functional integral is taken over all
field configurations which satisfy the periodic boundary condition
\begin{equation}
  \phi (t_i-i\beta ,\vex ) = \phi (t_i,\vex ) \ .
\end{equation}
One convenient choice of the complex time path consists of three
segments, $C=C_1\cup C_2\cup C_3$, where $C_1$ runs along the real
axis from $t_i$ to some $t_f\gg t_i$, $C_2$ runs backwards along
the time axis from $t_f$ to $t_i$, and finally $C_3$ runs parallel
to the imaginary axis from $t_i$ to $t_i-i\beta$ (Fig. \ref{fig:tpath2.eps}).
\myfig{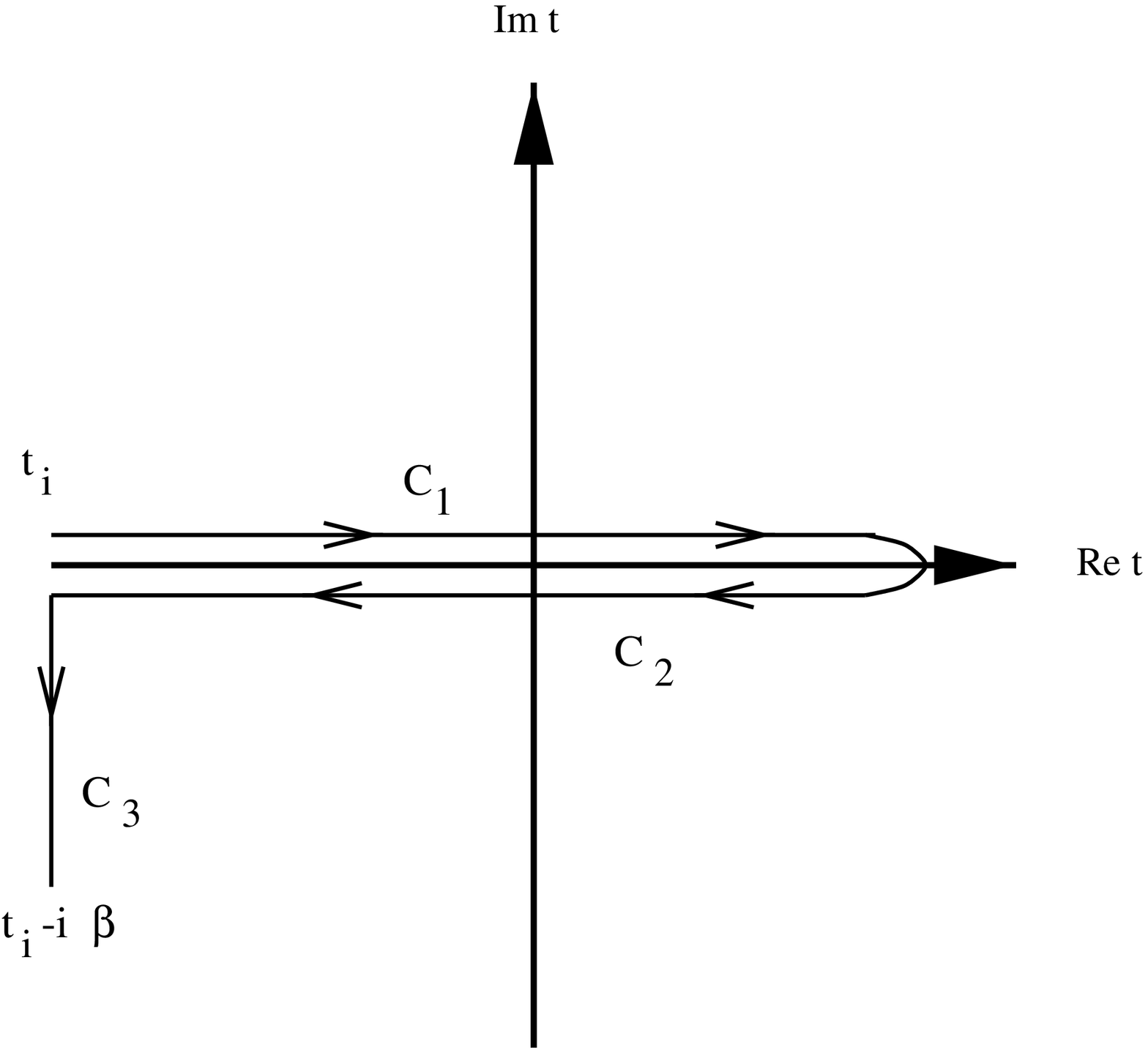}{4}{The time contour for FTQFT.}

\noindent
{}From the
generating functional, one can calculate the thermal Green
function
\begin{equation}
    iD_C(x-x')= \bra T_C \phi (x)\phi (x') \ket \ ,
\end{equation}
where time ordering has been promoted to path ordering $T_C$ along
the complex time path $C$. Equivalently, one can rewrite the
thermal Green function in terms of a $3\times 3$ matrix
$(D^{rs})_{r,s=1,2,3}$ with components
\begin{equation}
   D^{rs}(t-t') = D_C (t_r - t_s)
\end{equation}
where $t=t_r,\ t'=t_s$ if $t\in C_r$ and $t'\in C_s$ along
$C=C_1\cup C_2\cup C_3$.
Furthermore, 
if one takes $t_i,t_f\rightarrow -\infty, +\infty$ in an
appropriate manner, the contributions involving the segment $C_3$
decouple from the rest\footnote{Strictly speaking, in order to
take the energy eigenvalues correctly into account in the
amplitudes of the interacting thermal theory, one needs to also
take into account contributions from the vertical part of the
contour \cite{Kraus:2002iv}. However, this subtlety will not
affect the remaining discussion in this paper. We thank Per Kraus
for pointing this out.}. The matrix $(D^{rs})$ then reduces to a
$2\times 2$ matrix, but the temperature dependence remains, as the
components depend on the distribution function of the thermal
background. The contour $C$ reduces to the Schwinger-Keldysh
contour \cite{Schwinger:1961qe,Keldysh:1964ud} $C_1\cup C_2$. The
propagator is reproduced by breaking up the field $\phi$ and the
source $J$ into two-component vectors, \bea
  \phi &=& (\phi_1,\phi_2) \ \ {\rm with}\
  \phi_r(x)=\phi(t_r,\vex),\ t_r\in C_r \nonumber \\
  J &=& (J_1,J_2) \ \ {\rm with }\ J_r(x)=J(t_r,\vex),\ t_r\in C_r
  \ .
\eea The generating functional (\ref{ZFT}) then reduces to a form
\begin{equation}
  Z[J_1,J_2] = \int \Dcal \phi_1\Dcal \phi_2~\exp
  \left\{ i\int^{\infty}_{-\infty} d^4x~[\phi_r (D^{-1})^{rs} \phi_s
  + \Lcal_{int} (\phi_1)-\Lcal_{int} (\phi_2) + J_r\phi_r] \right\} \
\label{ZFT2}
\end{equation}
where $\Lcal_{int}$ is the interaction part of the Lagrangian. In
particular, the diagonal components of the $2\times 2$ propagator
$D^{rs}$ have the momentum space representation \bea iD^{11}(k)
&=& \frac{i}{k^2-m^2+i\epsilon}+2\pi \delta (k^2-m^2) n_T(k_0)
\nonumber \\ iD^{22}(k) &=& \frac{-i}{k^2-m^2-i\epsilon}+2\pi
\delta (k^2-m^2) n_T(k_0)
\eea where $n_T(k_0)$ is essentially the thermal distribution
function. It is then evident that in addition to the physical
field $\phi_1$, the theory contains another degree of freedom
$\phi_2$, called the thermal ghost, which propagates backwards in
time. The two fields $\phi_{1,2}$ are coupled together only by the
off-diagonal elements $D^{12,21}$ of the propagator. One is
interested in correlation functions of $\phi_1$ only. Furthermore,
it can be shown that in the zero temperature limit $\beta
\rightarrow \infty$ the off-diagonal elements of the propagator
vanish, $D^{12,21}\rightarrow 0$, so that the thermal ghost
decouples from the physical degree of freedom. Hence at zero
temperature one can ignore the thermal ghost and the theory
reduces back to the usual form involving only the physical degree
of freedom. But at any finite $T$, both fields make physical
contributions.

\paragraph{The orbifold case.} In the above example, the Fock
spaces associated with the physical field and thermal ghost are
the positive and negative energy sectors $\Hcal^+$ and $\Hcal^-$.
At zero temperature, before removing the thermal ghost, the
generating functional (\ref{ZFT2}) is symmetric under $\Zmath_2$
reflection which reverses the direction of time. This is precisely
what we need as a starting point for QFT on the covering space of
$\orbid$. We will also start with a path integral involving the
Schwinger-Keldysh contour as the complex time path. Then, as
before, we break up the field $\phi$ as a two-component vector
$\phi =(\phi_+,\phi_-)$ where $\phi_+$ and $\phi_-$ involve times
at the forward and backward running segments of the Schwinger-Keldysh contour.
The path integral can then be rewritten as
\begin{equation}
  Z = \int \Dcal \phi_+ \Dcal \phi_-~\exp \left\{
  i \int^{\infty}_{-\infty}dt \int d^d\vex~[\Lcal (\phi_+)-\Lcal (\phi_-)]
  \right\} \ ,
\label{Zorbi}
\end{equation}
where $\Lcal$ is the (for example) scalar field Lagrangian
\begin{equation}
 \Lcal (\phi) = \half (\pat \phi )^2 -\half m^2 \phi^2
 -V_{int}(\phi) \ .
\end{equation}
We have made physical input here by the choice of
propagator for $\{\phi_+,\phi_-\}$. On the covering space, our picture
is that the field $\phi_+$
propagates forward and its copy field $\phi_-$ propagates backward in time,
decoupled from each other for $t\neq 0$. Hence the propagator is diagonal
in $\phi_+,\phi_-$. We then choose the $t=0$ hypersurface as the time
slice where we define initial conditions\footnote{Hence we are choosing
the fundamental domain to be that of Figures \ref{fig:pic1.eps} b) and
\ref{fig:figg4.ps}. While this is the most convenient choice
for our QFT construction, other choices would also be possible.}.
More precisely, we could
consider the time evolution of $\phi_+$ from $t<0$ up to a specified
profile at $t=0$ and then forward to $t>0$, and the reverse for
$\phi_-$. The orbifold identification
then calls us to identify the fields and time evolutions (elaborated
further below). However, note first a subtlety in defining the initial condition.
At $t=0$ the orbifold identification is $(0,x)\sim (0,-x)$, hence
the profiles of $\phi_+$ and $\phi_-$ must become symmetric at $t=0$.
The most natural initial  condition is to set the profiles to be equal
at $t=0$. Thus
our initial condition is
\begin{equation}
x>0:\ \ \ \phi_+(0,x)=\phi_+(0,-x)=\phi_-(0,x)=\phi_-(0,-x)=\phi_0 (x)
\end{equation}
where $\phi_0(x)$ is the specified initial profile on $x>0$. This can be satisfied
as follows. Decompose the fields $\phi_\pm$ into symmetric and
antisymmetric parts under $x\mapsto -x$:
\bea
&& \phi_\pm(t,x)=\phi_{\pm,S}(t,x) + \phi_{\pm,A}(t,x) \ , \nonumber \\
&& \phi_{\pm,S}(t,x)=\half ( \phi_\pm(t,x)+\phi_\pm(t,-x)) \nonumber \\
&& \phi_{\pm,A}(t,x)=\half ( \phi_\pm(t,x)-\phi_\pm(t,-x)) \ .
\eea
The initial condition can be satisfied if the antisymmetric parts
$\phi_{\pm,A}$ decay to strictly zero sufficiently
rapidly as $t\rightarrow 0$ (and the symmetric parts become equal).
There is a subtlety here in what exactly should be
meant by ``sufficiently rapid,'' and we will
comment on it further below.

While the above serves as a starting point for our construction of the
theory on the covering space, we must also take into
account the identification which is part of the orbifold
construction. We already discussed this in the context of Fock space
states, and can now do it more explicitly. Let us leave
the path integral formalism and return
back to the canonical quantization prescription. In the remainder
of this section, we focus only on the free field part of the
Lagrangian. This is sufficient for the construction of the
invariant Fock space, and for the improved invariant version of the
back-reaction calculation which will replace that of Section \ref{sec:tradcalc}.
We will present a
tentative discussion of interacting theory and the S-matrix in
Section \ref{sec:smat}.

First, we quantize the field operators $\phi_\pm$.
While $\phi_+$ has the standard free field mode expansion
\begin{equation}
  \phi_+ (t,\vex ) = \int \frac{d^d\vek}{(2\pi )^d}~\frac{1}{\sqrt{2\omega_k}}
  \left\{ a_{\vek}~e^{-i\omega_{\vek} t + i\vek \cdot \vex }
  +  a^{\dagger}_{\vek}~e^{+i\omega_{\vek} t - i\vek \cdot \vex }
  \right\} \ ,
\end{equation}
for the operator $\phi_-$ we write the mode expansion as
\begin{equation}
 \phi_- (t,\vex ) = \int \frac{d^d\vek}{(2\pi )^d}~\frac{1}{\sqrt{2\omega_k}}
  \left\{ \atilde_{\vek}~e^{+i\omega_{\vek} t - i\vek \cdot \vex }
  +  \atilde^{\dagger}_{\vek}~e^{-i\omega_{\vek} t + i\vek \cdot \vex }
  \right\} \ .
\end{equation}
The initial condition at $t=0$ and the required rapid decay of the
antisymmetric parts of $\phi_\pm$ create subtleties, but the above
mode expansions are valid sufficiently far from the $t=0$ slice.
Since the field $\phi_-$ is decoupled from $\phi_+$, violations of
causality do not arise.

We can now present an improved (completely
$\Zmath_2$ invariant) version of the calculation of
the vacuum expectation value of the stress tensor.
Since the initial condition
creates subtleties near $t=0$, we first assume $|t|>0$ so that
we can trust the mode expansions. Then, simply
\bea
 {}_{inv}\bra 0| T^{inv}_{\mu\nu}(x)|0\ket_{inv}
 &=& \half \left( \bra 0| , \bra \tilde{0}| \right)
\left( \begin{array}{cc}
  T^+_{\mu\nu} (x) & \mbox{} \\ \mbox{} & T^-_{\mu\nu} (-x)
  \end{array} \right)
  \left( \begin{array}{c} | 0\ket \\
     |\tilde{0}\ket \end{array} \right) \nonumber \\
     \mbox{} &=& \lim_{x'\rightarrow x}
     \half \left\{ \bra 0 |\pat_\mu \phi_+ (x)
     \pat_\nu \phi_+ (x') |0\ket +
\bra \tilde{0} |\pat_\mu \phi_- (-x)
     \pat_\nu \phi_- (-x') |\tilde{0}\ket \right\} \nonumber \\
\mbox{} &=& \lim_{x'\rightarrow x}  \half\left\{
\frac{1}{(x-x')^2} +\frac{1}{(x-x')^2}\right\}
= \lim_{x'\rightarrow x} \frac{1}{(x-x')^2} \ .
\eea
This is again just the usual vacuum divergence. The renormalized
expectation value of $T_{inv}$ would then be equal to zero.
There are two  main
differences with the previous calculation of Section 2:
i) Everything is $\Zmath_2$ invariant, including the vacuum state.
ii) Essentially, $\phi(-u)$ is now replaced by $\phi_-$. But
$\phi_\pm$ are decoupled, so there are no ``$\bra 0 |\phi_+\phi_-|0\ket$"
cross contractions, as were depicted in Fig. \ref{fig:figgg5.ps}.


Near the initial slice $t=0$ the situation is more subtle. As noted
previously, the antisymmetric
part of the fields must die off sufficiently rapidly. Such
behavior will alter the mode expansion of the fields. If
we insist on trusting the mode expansion everywhere such that $t\neq0$, then
we must switch off the antisymmetric parts abruptly with step functions:
\begin{equation}
  \phi_{+,A} (t,\vex ) = (1-\theta(t))~f_+(\vex ) \ ; \
  \phi_{-,A} (-t,-\vex ) = (1-\theta(-t))~f_-(-\vex)  \ ,
\end{equation}
separating out the time dependence.
But then the $tt$ component of the invariant stress tensor will have
a $\delta^2 (t)$ singularity and the $tx^i$ components a $\delta (t)$
singularity at the initial slice $t=0$. However, if we interpret
the $\orbid$ orbifold as a toy model of cosmology, then the $t=0$ slice
plays the role of the initial singularity.
Having a divergent stress tensor
at the $t=0$ slice is then natural in such a cosmological
interpretation --- it could represent the necessity for appropriate
boundary conditions.

However, it is not clear how seriously we should take the initial
condition leading to the
divergence, as we did not derive it from a low-energy limit of a
string calculation. The low-energy limit the first quantized string
analysis of Section \ref{sec:stringcalc} yields no such thing.
Moreover, the existence of the twisted sector, localized
at the orbifold singularity, is related to more involved question of
whether there is a blow-up mode and whether the singular geometry
really is the actual geometry on which to consider the QFT.
We leave these questions for future work.
The main point that we would like to stress here is that the
stress tensor does not have a divergence from back reaction as the naive analysis would have suggested.
That kind of a singularity would have been a signal of a more
serious instability.

Note also that although the covering space first appeared
to have CTCs everywhere, in the
reformulation of QFT it is also apparent that nothing actually
propagates along a CTC. Instead of a single quantum propagating
around and around in a CTC, there is a quantum and its copy which
propagate in opposite directions. More precisely\footnote{We thank
Simon Ross for the following elegant argument.}, if a particle on
the covering space starts out at $t<0$ with a future-directed
tangent vector, when it reaches the image point at $t>0$, its
future-directed tangent vector corresponds to a past-directed
tangent vector at the starting point. So the future-going particle
at $t>0$ is then identified with a past-going particle at the
initial point $t<0$. Thus the particle cannot loop around the CTC
on the covering space, since its initial condition is not
repeated. What the choronology protection conjecture is meant to
forbid is closed paths that a particle can follow and return to
its initial condition, this is the essence of a time-machine. This
is not possible in the reformulated QFT, therefore it is not a
surprise that the instabilities associated with looping
around CTCs also do not arise.

\section{Time Evolution and Local S-matrix on the Fundamental Domain}
\label{sec:smat}

In all previous discussions, we limited the analysis to what
corresponds to free field theory in the low-energy limit. What
then of the interacting theory? Given that the orbifold is
globally time-nonorientable, is there any way of defining
S-matrices at least locally, for example away from the $t=0$ axis
in the above choice of the fundamental domain? A full
analysis of these questions is beyond the scope of the present paper, but we
can make some tentative comments and proposals.
Let us first illustrate the problem with a simple figure.
\myfig{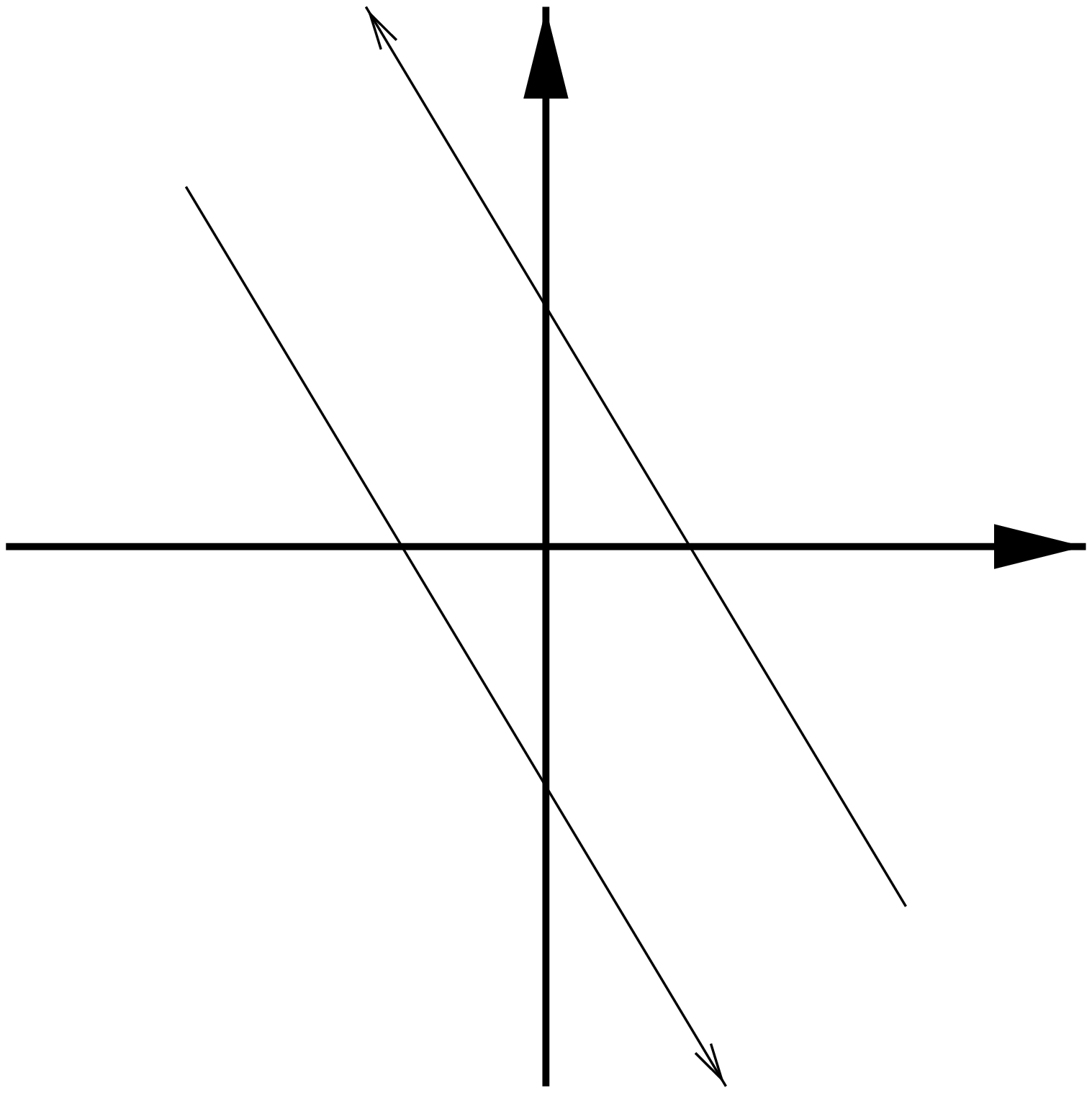}{5}{Example of a point particle and its image propagating
on the covering space.}
Figure \ref{fig:apu.eps} depicts a point particle and its image propagating in
opposite time and space directions on the covering space. As
discussed in Section \ref{sec:qftnew}, both trajectories will be identified by
the $\Zmath_2$ reflection, resulting in a single trajectory for a
point particle on the fundamental domain. The details depend on
the choice of the fundamental domain.

Consider first choosing the
right half-space as the fundamental domain, and drawing the
corresponding "pocket", as depicted in Figure \ref{fig:fig9.eps}
(see also Figs. \ref{fig:pic1.eps}(a) and \ref{fig:figgg3.ps}).
\myfig{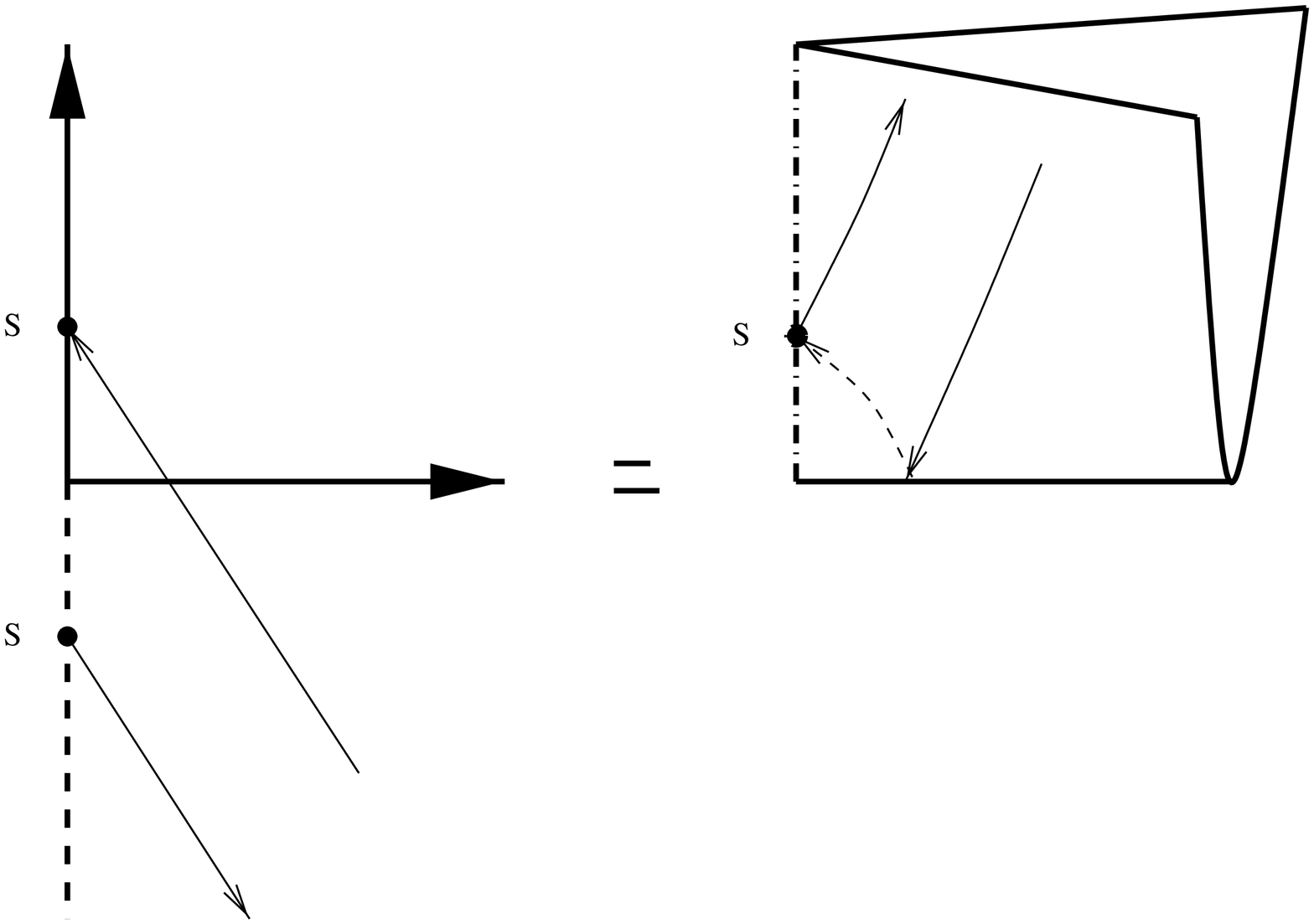}{7}{Point particle propagating on the fundamental domain.
The left part of the figure depicts the
fundamental domain, with the dashed negative time axis ($x=0$) identified
with the positive time axis. For example, the points S marked
with a black dot are identified. The identification results in the
pocket shown on the right. The time orientation breaks down on the
time axis. The part of the trajectory drawn with
solid lines depicts propagation on the front fold
of the pocket, while the dashed line depicts propagation on the rear fold. The
point S is on dotted-dashed line, where the time direction
becomes ill-defined.}
Alternatively, we can choose the upper half-space as the
fundamental domain, and identify the negative $x$-axis with the
positive $x$-axis. The result is depicted in Figure \ref{fig:fig10.eps}
(see also Figs. \ref{fig:pic1.eps}(b) and \ref{fig:figg4.ps}).
\myfig{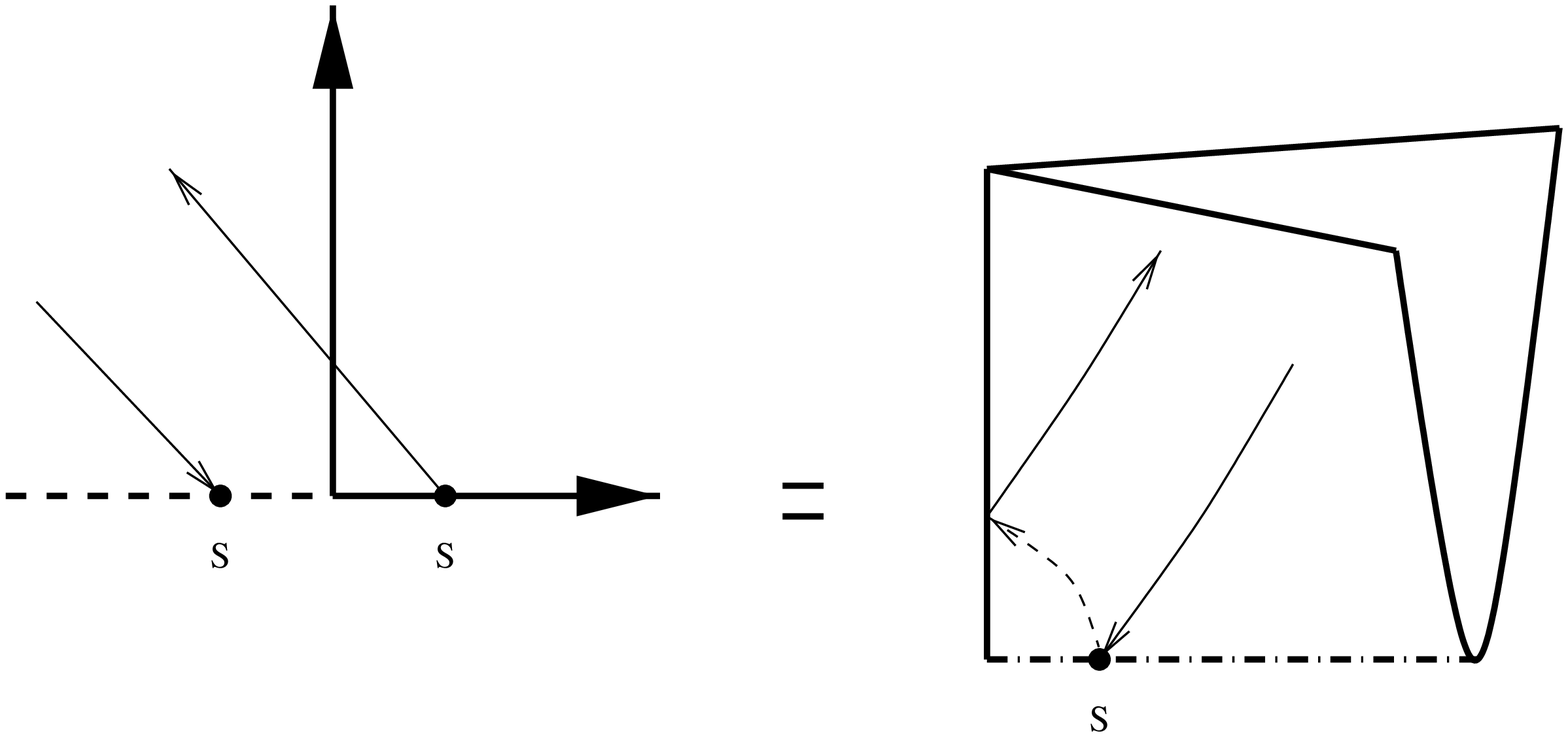}{7}{Point particle propagating on the fundamental domain.
On the left figure, the points marked with S are identified.
On the right figure, the solid lines depict propagation on the front fold
of the pocket, the dashed line depicts propagation on the rear fold. The
point marked with S is on the dotted-dashed line, where the
time direction becomes ill-defined.}

As discussed in Section \ref{sec:oldorbid}, a possible choice for the
time-arrow on the fundamental domain of Figure \ref{fig:fig9.eps} is to
let it point from
the lower right quadrant to the upper right quadrant, while
becoming ambiguous on the $x=0$ axis. On the pocket, time would
thus flow down the front fold and continue upwards on the rear
fold. Similarly, on the fundamental domain of
Figure \ref{fig:fig10.eps} one can choose the arrow of time to point
upwards on the upper half-space, with the
$t=0$ axis as the origin of time. On the pocket, time would then
flow upwards on both sides of the fold. However, then the
trajectories depicted on the figures would seem to violate causality.
On Figure \ref{fig:fig9.eps}, if we choose a constant time
slice far up on the front fold, the trajectory will cross it
twice. First it crosses the slice on its way down along the front
fold, then it continues to the other side but returns back to the
front slice and crosses the slice again. On Figure \ref{fig:fig10.eps}, the
trajectory would propagate first backwards in time towards $t=0$,
then propagate forward in time on the rear fold and again on the
front fold. Both interpretations are troublesome.

However, we can improve the situation a bit. In Section \ref{sec:qftnew},
we identified forward time evolution of a particle
with backward time evolution of its image on the covering space.
We start the forward evolution from $t=-\infty$ and the backward
evolution from $t=\infty$. The time evolution continues without
problems until we reach the dividing line between the
two half-spaces, depending on the choice of the fundamental
domain. That is, if we choose the right half-space as the
fundamental domain, we can follow the time evolution until the
particle and its image reach $x=0$. If we choose the upper
half-space as the fundamental domain, the time evolution can be
followed up to $t=0$. Similarly, just after crossing the dividing
line, we can again follow the time evolution onwards. For example,
in the latter case we can continue from $t=0+\epsilon$ the
forward time evolution to $t=\infty$ and backward evolution to
$t=-\infty$. The problem is if and how it is possible to continue
the evolution across the dividing line.

If we choose the upper half-space as the fundamental domain, it is
simple to give a more formal definition.
In the Heisenberg picture we define the invariant time
evolution operator from $t_0$ to $t_1>t_0$ on the covering space to be
\bea
   U_{inv}(t_0,t_1) &=& T\left\{\exp \left[ -i\int^{t_1}_{t_0}
   dt~H_{inv}(t)\right] \right\} \nonumber \\
   \mbox{} &=& \left( \begin{array}{cc} T\{\exp [-i\int^{t_1}_{t_0}
   dt~H^+]\} & 0 \\
0 & \tilde{T}\{\exp [-i\int^{-t_1}_{-t_0} dt~H^-]\} \end{array} \right)
\nonumber \\
\mbox{} &=&
\left( \begin{array}{cc} U^+(t_0,t_1) & 0  \\
0 & U^-(-t_0,-t_1) \end{array} \right) \ ,
\eea
where $\tilde{T}$ denotes anti-time ordering. This is unambiguously defined
if both $t_0,t_1 <0$ or both $t_0,t_1>0$. Problems arise
when $t_0<0$ and $t_1>0$.

Let us see what this means for the point particles in the figures.
In Figure \ref{fig:apu.eps}, we launch the particle and its image from $t=-\infty$
and $t=+\infty$, and then propagate them using $U(-\infty ,t)$
towards the $x$-axis, {\em i.e.} up to $t=0-\epsilon$. This gives
the lower half and the upper half of the forward and backward
trajectories of Fig. \ref{fig:apu.eps}. They are identified on the
fundamental domain,
so  this gives the ``downward" trajectory on the left part
in Figure \ref{fig:fig10.eps} all the way to the marked point S, and the downward
trajectory to the point S on the front fold of the pocket in
Figure \ref{fig:fig10.eps}. Similarly, we could propagate forward and backward
from $t=0+\epsilon$, giving the other halves of the trajectories
in Figure \ref{fig:apu.eps}. In Figure \ref{fig:fig10.eps}, the resulting
trajectory is the ``upward"
one on the left diagram, and the upward trajectory starting on the rear
fold and continuing to the front fold of the pocket.

To summarize,
on the fundamental domain (the pocket), we can either choose the time
to point downwards, corresponding to Figure \ref{fig:pic1.eps} b) with
the arrows reversed, and consider time evolution up to a ``big
crunch" at $t=0$, or choose the time to point upwards and consider
time evolution forward from a ``big bang". The latter
corresponds to Figure \ref{fig:pic1.eps} b). Either way, the evolution breaks
down at $t=0$. But that is also the point where from Section \ref{sec:qftnew}
we know the stress tensor to diverge\footnote{Similar analysis, based
on the other two choices of the fundamental domain, Figure \ref{fig:pic1.eps}
a) and c),
are also possible. Then the time evolution would break down at $x=0$
(a) or at null cone (c).}.
The choice could be interpreted as the additional projection on degrees
of freedom, discussed in section 5., with the remaining number of
states being one half of those in Minkowski space, in analogue
with the Euclidean $\Zmath_2$ orbifold.

Consider then an interacting field theory. We could adopt the
interaction picture, and define the time evolution operator as
\bea
 &&  U_{I,inv} (t_0,t_1) = \nonumber \\
 &&
\left( \begin{array}{cc} T\{\exp [+i\int^{t_1}_{t_0}
   dtd^d\vex~{\cal L}_I (\phi_+(t,\vex))]\} & 0 \\
0 & T\{\exp [-i\int^{t_1}_{t_0} dt' d^d\vex~
{\cal L}_I(\phi_-(t',\vex))]\} \end{array} \right)
\eea
where ${\cal L}_I$ is the interaction part of the Lagrangian.
We could consider this as the local S-matrix. Hence, as long
as we stay away from the singular $t=0$ hypersurface, the S-matrix
has the same properties as that of an ordinary field theory on
Minkowski space.

\bigskip

\noindent
{\large {\bf Acknowledgment}}

\bigskip

We would like to thank Vijay Balasubramanian, Fawad Hassan, and
Asad Naqvi for the collaboration which lead us to this
investigation, and many discussions on the interpretational
issues. We also thank Per Kraus for helpful discussions and useful
encouragement, and Simon Ross for useful discussions on
time-nonorientability and time travel. EK-V was in part supported
by the Academy of Finland, and thanks the University of California
at Los Angeles and University of Pennsylvania for hospitality at
different stages of this work. RGL was supported by US DOE under
contract DE-FG02-91ER40677 and thanks the Helsinki Institute of
Physics for hospitality. ES would like to thank A. Lawrence, S.
Kachru, A. Knutson and R. Bryant for helpful conversations.
Finally, we thank the Aspen Center for Physics for hospitality, and
the participants of the workshop "Time and String Theory" for
useful discussions, as we were re-editing this paper.

\bigskip
\begin{appendix}
\section{Appendix}
\newcommand{\RR}{{\Rmath}}
\newcommand{\ZZ}{{\Zmath}}
\newcommand{\pa}{\partial}

In this appendix, we will provide the details of complementary calculations
using oscillator methods. There are several subtleties that are not regularly
seen in the usual backgrounds.
We will use the notation where $\tilde k$ is the image of $k$ under the orbifold.
\bea
\tilde k_o=-k_o\\
\tilde k_u=+k_u
\eea
We want to evaluate
\begin{equation}
T(\tau)=\frac{1}{2}\tr_{U+T} \left[\left( 1+\hat g\right)
\int d^2z V(z,\bar z)\  q^{L_o-a}\bar q^{\tilde L_o-\tilde a}\right]
\end{equation}
It is important to note that in this formalism, obtained by sewing the cylinder
into a torus, there are zero modes in the $U$ sectors of the trace, but not
in the twisted $T$ sectors.
The massless vertex operator is of the form
\begin{equation}
V(z,\bar z)=\frac{2g_{str}}{\alpha'}:\partial X^\mu\bar\partial
X^\nu\frac{1}{2}\left( e^{ik\cdot X(z,\bar z)}
+ e^{i\tilde k\cdot X(z,\bar z)}\right):
\end{equation}

The non-zero mode portion of this expression can be evaluated using
coherent state methods.
For each oscillator $\alpha_n^\mu$ ($n>0$) we introduce a coherent-state
basis $|\rho_{n,\mu})$
and write the trace as a $\rho$-integral. If the $\partial X^\mu$ does not
contribute an oscillator, we find (for each $n>0$ and $\mu$)
\begin{equation}
\int \frac{d^2\rho}{\pi}\ e^{-|\rho|^2} e^{\alpha' k_\mu^2/4n}(\rho|
e^{\sqrt{\alpha'/2}k_\mu \alpha_{-n}^\mu z^n/n}
e^{-\sqrt{\alpha'/2}k_\mu \alpha_{n}^\mu z^{-n}/n}|q^n\rho)
\end{equation}
 for the $1$-insertion, while for the $\hat g$-insertion, we get
\begin{equation}
\int \frac{d^2\rho}{\pi}\ e^{-|\rho|^2} e^{\alpha' k_\mu^2/4n}(\rho|
e^{\sqrt{\alpha'/2}\tilde k_\mu \alpha_{-n}^\mu z^n/n}
e^{-\sqrt{\alpha'/2}\tilde k_\mu \alpha_{n}^\mu z^{-n}/n}|-q^n\rho)
\end{equation}
This is a standard integral whose evaluation can be
found in \cite{Green:1987sp}. The result is
\begin{equation}
\frac{1}{1\mp q^n} e^{\mp \alpha'k_\mu k^\mu \frac{q^n}{2n(1\mp q^n)}}
\end{equation}
again for each $n>0$ and $\mu$. For the $\tilde\alpha$ oscillators, we will
get the same result, with $q$ replaced by $\bar q$. Now by simple re-ordering of
sums and appropriate\footnote{In particular, there is a factor of 2 which must
be absorbed by the (implicit) regulator in the first equation. This can be seen,
for example, as a requirement of modular invariance.} renormalization, we
may compute:
\begin{equation}
\prod_{n\in\ZZ^+}e^{\frac{1}{n}\frac{q^n}{1+q^n}} =
\frac{\theta_2(\tau)}{q^{1/8}},\ \ \ \ \ \
\prod_{n\in\ZZ^+-1/2}e^{\frac{1}{n}\frac{q^n}{\pm 1+q^n}} = \theta_{3,4}(\tau)
\end{equation}

On the other hand, the $\partial X^\mu$ might contribute an oscillator. Then, we
have a new matrix element
\begin{equation}
\sum_{m>0}\int \frac{d^2\rho}{\pi}\ e^{-|\rho|^2} (\rho|
e^{\sqrt{\alpha'/2}k_\mu \alpha_{-n}^\mu z^n/n}
\left[z^{-m-1}\alpha_m^\mu+z^{m-1}\alpha_{-m}^\mu\right]
e^{-\sqrt{\alpha'/2}k_\mu \alpha_{n}^\mu z^{-n}/n}|q^n\rho)
\end{equation}
When $m=n$, we find, recalling that
$[\alpha_m,e^{a\alpha_{-m}}]=mae^{a\alpha_{-m}}$ and
$|\rho_n)=e^{\rho \alpha_{-n}/\sqrt{n}}|0\rangle$
\bea
\frac{\sqrt{m}}{z}\int \frac{d^2\rho}{\pi}\ e^{-|\rho|^2}
\left[z^{-m}q^m\rho+z^{m}\bar\rho\right]
(\rho| e^{k_\mu \alpha_{-n}^\mu z^n/n}
e^{-k_\mu \alpha_{n}^\mu z^{-n}/n}|q^n\rho)\\
=\frac{\sqrt{m}}{z}\int \frac{d^2\rho}{\pi}\ e^{-(1-q^m)|\rho|^2}
\left[z^{-m}q^m\rho+z^{m}\bar\rho\right]
e^{k_\mu(z^m\bar\rho-z^{-m}q^m\rho)/\sqrt{m}}
\eea
It is straightforward to show that this vanishes. Thus only the zero mode
part of the $\partial X^\mu$ factors contribute. As a corollary then, only
the untwisted sector will contribute to the massless tadpole in the Lorentzian
orbifold.

It remains to evaluate the zero modes. These are
\bea
U,1:&&\prod \int \frac{dp}{2\pi}\langle p| \hat P^\mu\hat P^\nu
e^{-\pi\alpha'\tau_2 \hat P^2}e^{(\alpha'/2) k\hat P\ln|z|^2}|p+k\rangle\
|z|^{-\alpha'k_o^2/2}= \frac{\eta^{\mu\nu}}{2\pi\alpha'\tau_2}\prod_o
\frac{\delta(\sqrt{\alpha'}k_o)}{\sqrt{\tau_2}}\nonumber\\
U,\hat g:&&\prod \int \frac{dp}{2\pi}\langle \tilde p| \hat P^\mu\hat P^\nu
e^{-\pi\alpha'\tau_2 \hat P^2}e^{(\alpha'/2) k\hat P\ln|z|^2}|p+k\rangle\
|z|^{-\alpha'k_o^2/2}=\\ &&=\prod_o\frac{e^{-\pi\tau_2\alpha' k_o^2/4}}{2}\times
\left\{\begin{matrix} k_\mu k_\nu/4& \mu,\nu\in o\cr \frac{1}{2\pi\alpha'\tau_2}&
\mu=\nu \in u\end{matrix}\right.\nonumber
\eea
each times a factor $-\left(\frac{\alpha'}{2}\right)^2\frac{1}{|z|^2}
\prod_u\frac{\delta(\sqrt{\alpha'}k_u)}{\sqrt{\tau_2}} $.
In the first case, this is multiplied by
$X_{1,1}= |\eta(\tau)|^{-24}$, while in the second,
we have $X_{1,0}=\prod_n | q^{-1}(1-q^n)^{d-23}(1+q^n)^{-(d+1)}|^2$.
Thus, if we go on-shell, we get
\begin{equation}
T^{\mu\nu}_0=-\left(\frac{g_{str}\alpha'}{2}\right)
\prod_u\frac{\delta(\sqrt{\alpha'}k_u)}{\sqrt{\tau_2}}
\frac{\eta^{\mu\nu}}{ 2\pi\alpha'}
\left( \prod_o \frac{\delta(\sqrt{\alpha'}k_o)}{\sqrt{\tau_2}}
X_{1,1}+2^{-(d+1)}\sum_{(g,h)\neq (1,1)}X_{g,h}\right)
\end{equation}
if $\mu,\nu$ are in the unorbifolded directions, while if they are in the
orbifolded directions
\begin{equation}\label{eq:nonmod}
T^{\mu\nu}_0=-\left(\frac{g_{str}\alpha'}{2}\right)
\prod_u\frac{\delta(\sqrt{\alpha'}k_u)}{\sqrt{\tau_2}}
\left( \frac{\eta^{\mu\nu}}{ 2\pi\alpha'}
\prod_o \frac{\delta(\sqrt{\alpha'}k_o)}{\sqrt{\tau_2}}
X_{1,1}+\frac{k_\mu k_\nu}{ 4} 2^{-(d+1)} X_{1,0}\right)
\end{equation}
This result is not modular invariant. However there is an ordering
ambiguity\footnote{This ambiguity does not appear for Euclidean orbifolds.}
in zero modes from the $(U,\hat{g})$ sector that we have not taken into account.
To see the problem, suppose we write the vertex operator as
\begin{equation}
V=\alpha \pa X^\mu\bar\pa X^\nu e^{ik.X}+\beta \pa X^\mu e^{ik.X}\bar\pa X^\nu+
e^{ik.X} \pa X^\mu\bar\pa X^\nu
\end{equation}
Then, the $k_\mu k_\nu /4$ in eq. (\ref{eq:nonmod}) is
multiplied by $\alpha+2\beta+\gamma$.
There is a modular invariant choice ($\alpha+\beta+\gamma=1$, $\beta=-1$)
for which the $kk$ terms cancel. (There is no other effect of this ordering issue.)
We then obtain (In the notation of
Section \ref{sec:stringcalc},
$X_{1,1}=\frac{\tau_2^{12}}{V_{26}}Z_{o,(1,1)}Z_u$)
\begin{equation}
T^{\mu\nu}_0=-\left(\frac{g_{str}}{4\pi \tau_2}\right)\eta^{\mu\nu}
\frac{ \delta^{(26)}(k)}{V_{26}}Z_{o,(1,1)}[\tau]Z_u[\tau]
\end{equation}
(for $\mu,\nu$ in the orbifold directions). This result agrees with the
result in Section \ref{sec:stringcalc} for the case of the Lorentzian orbifold.
\end{appendix}

\providecommand{\href}[2]{#2}\begingroup\raggedright\endgroup

\end{document}